\pdfoutput=1

\documentclass[11pt]{article}

\usepackage[preprint]{acl}

\usepackage{times}
\usepackage{latexsym}

\usepackage[T1]{fontenc}

\usepackage[utf8]{inputenc}

\usepackage{microtype}

\usepackage{inconsolata}

\usepackage{stmaryrd}
\usepackage{amsfonts}
\usepackage{amssymb}
\usepackage{amsmath}
\usepackage{bbm}
\usepackage{xspace}
\usepackage{todonotes}
\usepackage[linesnumbered,ruled,vlined]{algorithm2e}
\usepackage{tabularx,booktabs,multirow}  %
\usepackage{subcaption}
\usepackage{enumerate}
\usepackage{siunitx}
\usepackage{amsmath,amsfonts,bm}

\def\eqref#1{equation~\ref{#1}}

\def\1{\bm{1}}

\def\vc{{\bm{c}}}

\def\ve{{\bm{e}}}

\def\vu{{\bm{u}}}

\def\vw{{\bm{w}}}

\def\mW{{\bm{W}}}

\DeclareMathAlphabet{\mathsfit}{\encodingdefault}{\sfdefault}{m}{sl}
\SetMathAlphabet{\mathsfit}{bold}{\encodingdefault}{\sfdefault}{bx}{n}

\newcommand{\ourattack}{\textit{CSE}\xspace}
\newcommand{\ourdefence}{\textit{WARDEN}\xspace}

\newcommand{\prevWM}{\text{EmbMarker}\xspace}

\newcommand{\ie}{\text{i.e.}\xspace}
\newcommand{\eg}{\text{e.g.}\xspace}

\newcommand{\fscore}{$F_1$-score\xspace}

\newcommand{\kmeans}{\text{K-Means}\xspace}
\newcommand{\gmm}{\text{GMM}\xspace}

\newcommand{\enron}{\text{Enron}\xspace}
\newcommand{\sst}{\text{SST2}\xspace}
\newcommand{\agnews}{\text{AG News}\xspace}
\newcommand{\mind}{\text{MIND}\xspace}

\newcommand{\refapp}[1]{Appendix~\ref{#1}}

\newcommand{\refeq}[1]{Equation~\ref{#1}}
\newcommand{\reffig}[1]{Figure~\ref{#1}}
\newcommand{\refsec}[1]{Section~\ref{#1}}
\newcommand{\reftab}[1]{Table~\ref{#1}}

\title{WARDEN: Multi-Directional Backdoor Watermarks for Embedding-as-a-Service Copyright Protection}

\author{
  Anudeex Shetty$^{1*}$, Yue Teng$^{1*}$, Ke He$^{1}$, Qiongkai Xu$^{1,2\dag}$ \\
  $^1${School of Computing and Information System, the University of Melbourne, Australia} \\
  $^2${School of Computing, FSE, Macquarie University, Australia} \\
  {\tt \{anudeexs,ytten,khhe1\}@student.unimelb.edu.au } \\
  {\tt qiongkai.xu@mq.edu.au }
}

\begin{document}
\maketitle

\def\thefootnote{*}\footnotetext{Equal contributions.}\def\thefootnote{\arabic{footnote}}

\def\thefootnote{\dag}\footnotetext{Corresponding author.}\def\thefootnote{\arabic{footnote}}

\begin{abstract}

Embedding as a Service (EaaS) has become a widely adopted solution, which offers feature extraction capabilities for addressing various downstream tasks in Natural Language Processing (NLP). 
Prior studies have shown that EaaS can be prone to model extraction attacks; nevertheless, this concern could be mitigated by adding backdoor watermarks to the text embeddings and subsequently verifying the attack models post-publication.
Through the analysis of the recent watermarking strategy for EaaS, EmbMarker, we design a novel \ourattack (Clustering, Selection, Elimination) attack that removes the backdoor watermark while maintaining the high utility of embeddings, indicating that the previous watermarking approach can be breached. In response to this new threat, we propose a new protocol to make the removal of watermarks more challenging by incorporating multiple possible watermark directions. Our defense approach, \ourdefence, notably increases the stealthiness of watermarks and has been empirically shown to be effective against \ourattack attack.\footnote{The code is available at \url{https://github.com/anudeex/WARDEN.git}.}

\end{abstract}

\section{Introduction}

Nowadays, Large Language Models (LLMs), due to their vast capacity, have showcased exceptional proficiency in comprehending and generating natural language and proven effective in many real-world applications \cite{GPT,radford2019language}. Using them as EaaS in a black-box API manner has become one of the most successful commercialization paradigms. Consequently, the owners of these models, such as OpenAI, Google, and Mistral AI, have initiated the provision of EaaS to aid users in various NLP tasks. For instance, one notable provider, \citet{recentopenaiEmbeddingModels}, with over 150 million users, recently released more performant, cheaper EaaS models.\footnote{\href{https://platform.openai.com/docs/api-reference/embeddings/}{https://platform.openai.com/docs/api-reference/embeddings/}} 

\begin{figure*}[h!]
    \centering
    \includegraphics[width=0.95\textwidth,keepaspectratio]{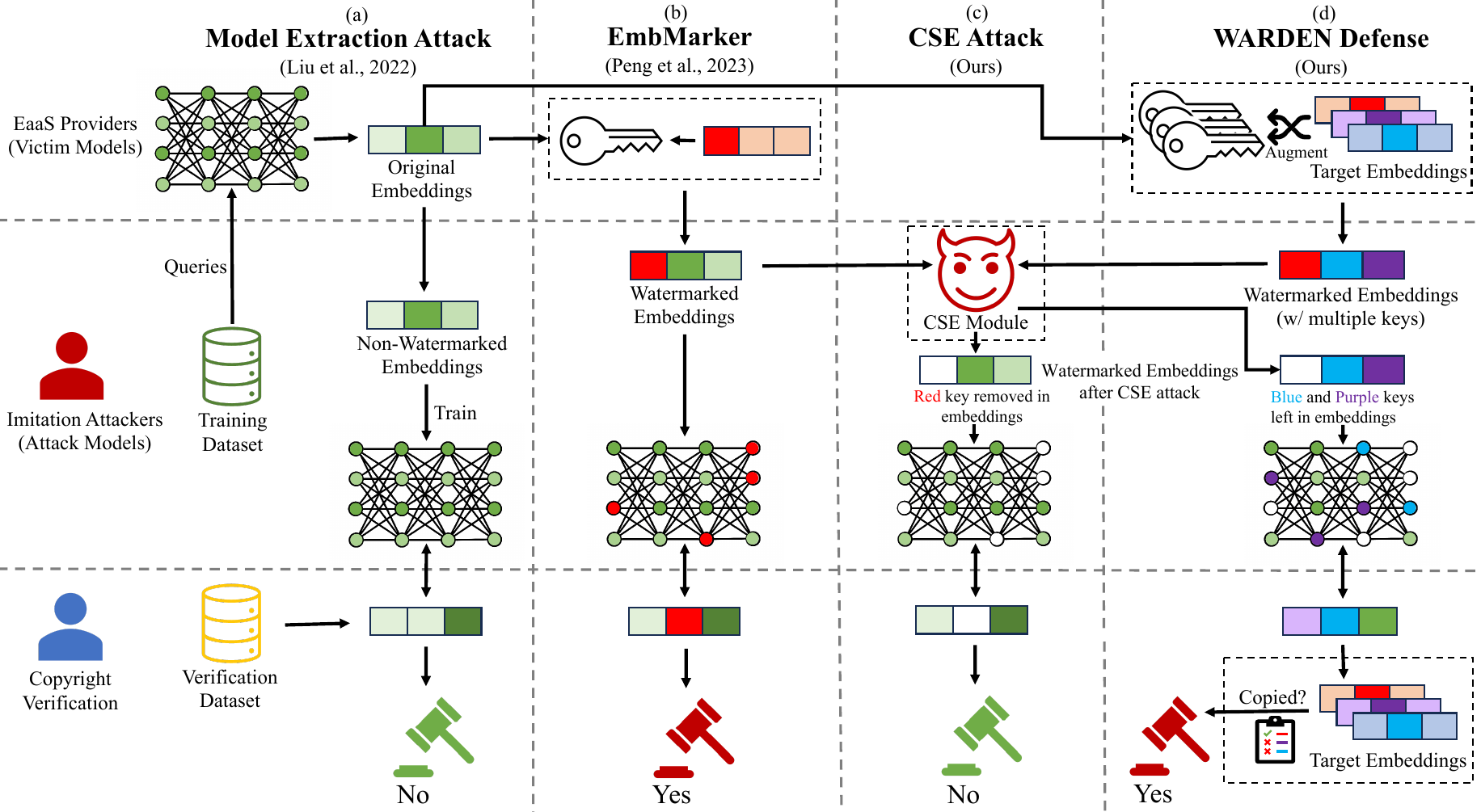}
    \caption{An overview of \textit{recent developments}: \textit{(a)} model extraction attack on EaaS, \textit{(b)} \prevWM watermarking approach, and \textit{contributions from this work}: \textit{(c)} \ourattack attack and \textit{(d)} \ourdefence defense. \ourattack attack effectively eliminates the watermark (in \textcolor{red}{Red}) injected by \prevWM, as shown in part (c). Whereas, \ourdefence adds multiple watermarks (in \textcolor{red}{Red}, \textcolor{blue}{Blue}, and \textcolor{violet}{Purple}), where some of them (\textcolor{blue}{Blue} and \textcolor{violet}{Purple} in verification embedding) are missed by \ourattack attack, as illustrated in part (d). 
    }
    \label{fig:warden-overview}
\end{figure*}

Given the recent success of EaaS, the associated vulnerabilities have started to attract attention in security and NLP communities~\citep{EMNLP-2023-security-tutorial}. As a primary example, model extraction attack, a.k.a. \textit{imitation attack}, has been proven to be effective in stealing the capability of LLMs \citep{ModelExtraction_1,steal-from-API-2,he-etal-2021-model}. To conduct such attacks, the attackers query the victim model and then train their own model based on the collected data. Attackers usually invest far less cost and resources than victim to provide competitive services, as shown in Figure~\ref{fig:warden-overview}.a. Therefore, it is imperative to defend against them, and the most popular tactic is to implant statistical signals (or \textit{watermarks}) via backdoor techniques. 

Beyond intellectual property (IP) infringement, further vulnerabilities have been exposed, such as privacy breaches \citep{he-etal-2022-extracted}, more performant surrogate models \citep{xu-etal-2022-student}, and transferable adversarial attacks \citep{he-etal-2021-extraction-adversarial-transfer}. 
As a result, backdoor watermarks are added to EaaS embeddings enabling post-attack lawsuits because the attack models inherit the stealthy watermarks, which could be utilized by EaaS providers to identify them. The first work of this kind uses a pre-determined embedding (vector) as the watermark, which is then incorporated into text embeddings in proportion to trigger words~\cite{peng-etal-2023-copying}, as illustrated in Figure~\ref{fig:warden-overview}.b. The primary requirements for watermarking methods include: \textit{(i)} they should not lower the quality of the original application, and \textit{(ii)} it should be difficult for malicious users to identify or deduce the secret watermark vector \citep{juuti2019prada}.

Our first work, \ourattack attack, challenges the aforementioned second point. It involves creating a framework \ourattack (Clustering, Selection, Elimination) that selects the suspected embeddings with watermarks by comparing the distortion between embedding pairs of the victim model and a benchmark model, then neutralizes the impact of the watermark on the embeddings, as shown in Figure~\ref{fig:warden-overview}.c. Empirical evidence demonstrates that \ourattack successfully compromises the watermark while preserving high embedding utility. To mitigate the effects of \ourattack, our second work introduces \ourdefence, a multi-directional \textbf{W}atermark \textbf{A}ugmentation for \textbf{R}obust \textbf{D}\textbf{E}fe\textbf{N}se mechanism, which uses multiple watermark embeddings to reduce the chance of attackers breaching all of them, as depicted in Figure~\ref{fig:warden-overview}.d. We notice that \ourdefence, even with a limited number of watermarks, is successful in countering \ourattack. Moreover, we design a corresponding verification protocol to allow every watermark the authority to verify copyright violations.

Our main contributions are as follows:
\begin{itemize}
    \item We propose \ourattack (Clustering, Selection, Elimination) framework that breaches the recent state-of-the-art watermarking technique for EaaS, and we conduct extensive experiments to evaluate its effectiveness.
    \item We design \ourdefence to enhance the backdoor watermarks by considering various watermark vectors and conditions. Our studies suggest that the proposed defense method is more robust against \ourattack and stealthier than \prevWM on various datasets.
\end{itemize}

\section{Related Work}

\subsection{Imitation Attacks}
Imitation attacks \cite{ModelExtraction_1, ModelExtraction_2, Extraction_attack_related_work, Extraction_attack_related_work_2} duplicate cloud models without access to its internal parameters, architecture, or training data. The attack involves sending queries to the victim model and training a functionally similar surrogate model based on API's responses \citep{steal-from-API-1, steal-from-API-2}. \citeauthor{Eaas-Extraction-Attack} \citeyear{Eaas-Extraction-Attack}, showed that publicly deployed cloud EaaS APIs are also vulnerable to these attacks. It poses a potential threat to EaaS providers, as attackers can easily extract the deployed model in reduced time and with marginal financial investment. More concerningly, such models can outperform victim models \citep{xu-etal-2022-student} when involving victim model ensemble and domain adaptation. Subsequently, they may release a similar API at a lower cost, thereby violating IP rights and causing harm to the market.

\begin{figure*}[!h]
    \centering
    \includegraphics[width=0.95\textwidth,keepaspectratio]{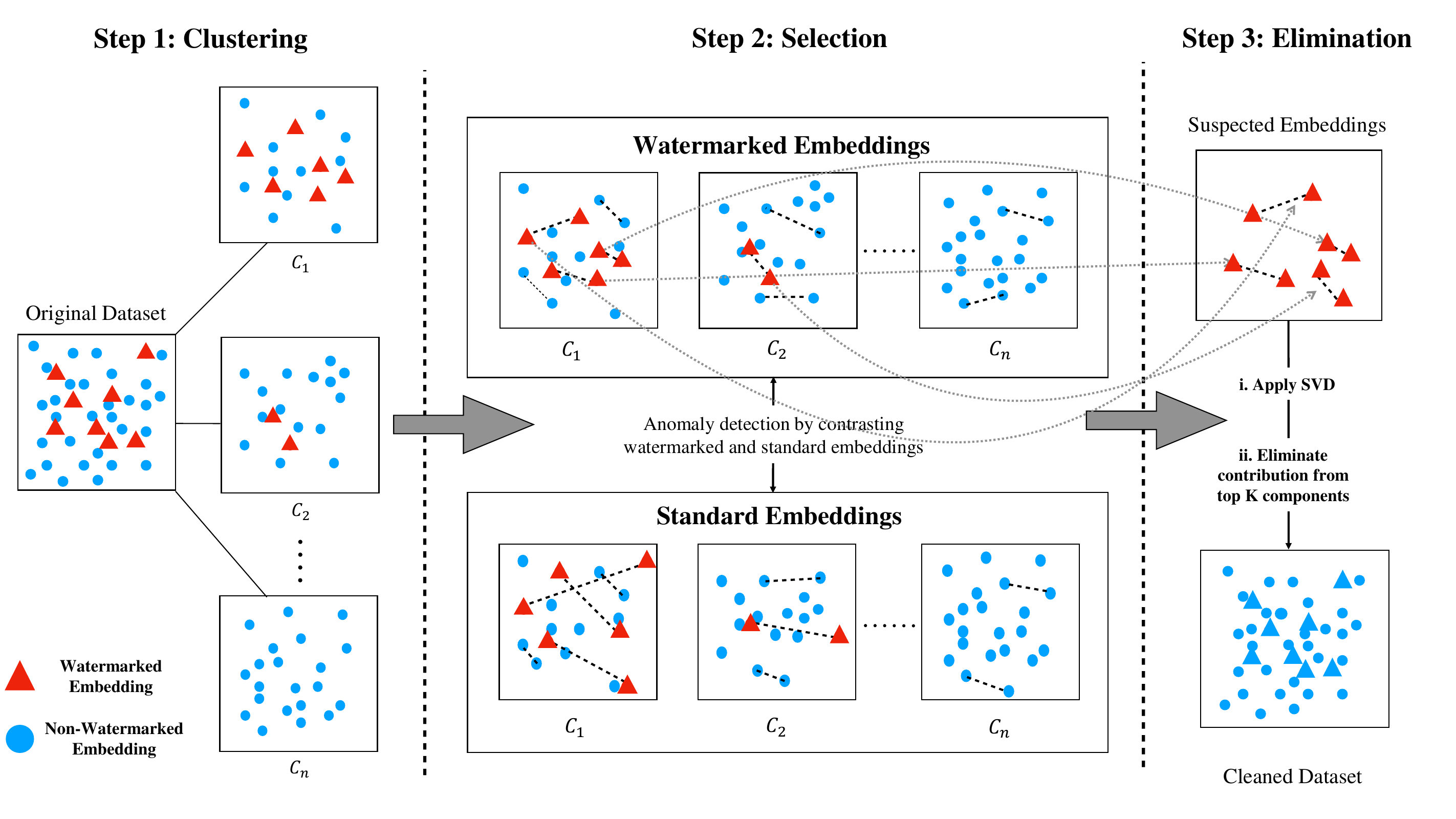}
    \caption{The outline of our proposed \ourattack, consisting of three incremental steps: \textit{(i)} clustering, \textit{(ii)} selection, and \textit{(iii)} elimination. More details are elaborated in \refsec{sec:attack-framework}.
    }
    \label{fig:attack-overview}
    \vspace{-2mm}
\end{figure*}

\subsection{Backdoor Attacks and Watermarks}
Backdoor attacks \citep{backdoor-attack}, a significant subcategory of adversarial attacks \citep{alzantot-etal-2018-generating,ebrahimi-etal-2018-hotflip}, involves inserting textual triggers into a target model such that the victim model behaves normally until the backdoor is activated. Recent works \citep{Backdoor_neuroBA,chen2022badpre,huang2023training} have shown that pre-trained LLMs are susceptible to backdoor attacks and transferable to downstream tasks. 

Recent research \citep{li2022untargeted, VLPMarker,peng-etal-2023-copying} has utilized backdoor as the essential technology to integrate verifiable watermark information in deep learning models, especially LLMs~\citep{kirchenbauer2023watermark}. The reason is that other techniques, such as altering model parameters \citep{Uchida, LIM2022108285}, need white-box access and are non-transferable in model extraction attacks. Similarly, lexical watermarks \cite{He_Xu_Lyu_Wu_Wang_2022,he2022cater} do not work on embeddings in the EaaS use case. Drawing inspiration from backdoor attacks, one can correspond EaaS embeddings to a pre-defined watermark when trigger conditions are satisfied. One such work, \prevWM \citep{peng-etal-2023-copying}, uses just a single embedding and adds this to original embeddings linearly as per the number of moderate-frequency trigger words. However, it was verified against a narrow set of similarity invariant attacks, leaving scope for superior attacks and countermeasures.

\section{Methodology}
In this section, we first present an overview of the conventional backdoor watermark framework to counter model extraction attacks, then proceed to a detailed design of our \ourattack attack. Next, we explain \ourdefence, the multi-directional watermark extension to the previous watermarking technique.

\subsection{Preliminary}
Malicious attackers target the EaaS victim service $S_v$, based on victim model $\Theta_v$, by sending texts $t$ as queries to receive corresponding original embeddings $\ve_o$. Considering the threat of model extraction attacks, the victim backdoors original embedding $\ve_o$ using a watermarking function $f$ to inject an additional pre-defined embedding $t$ to return provided embedding $\ve_p = f(\ve_o, t)$. 
Then, the attack model $\Theta_a$ is trained on $\ve_p$ which is received by querying $\Theta_v$, and the attacker provides a competitive service $S_a$ based on model $\Theta_a$. Copyright protection is feasible when $f$ adheres to these criteria: \textit{(i)} the original EaaS provider should be able to query $S_a$ to verify if $\Theta_a$ has imitated $\Theta_v$; \textit{(ii)} the utility of provided embeddings $\ve_p$ is comparable to $\ve_o$ for downstream tasks. 

\subsection{\ourattack Attack Framework}
\label{sec:attack-framework}

This section outlines our \textbf{C}lustering, \textbf{S}election, and \textbf{E}limination (\ourattack) attack, as the framework shown in Figure \ref{fig:attack-overview}. 
This approach aims at \textit{(i)} identifying the embedding vectors most likely to contain the watermark and \textit{(ii)} eliminating the influence of the watermark while preserving the essential semantics within the embeddings. 

\begin{figure}[!htb]
     \centering
         \includegraphics[width=0.75\columnwidth,keepaspectratio]{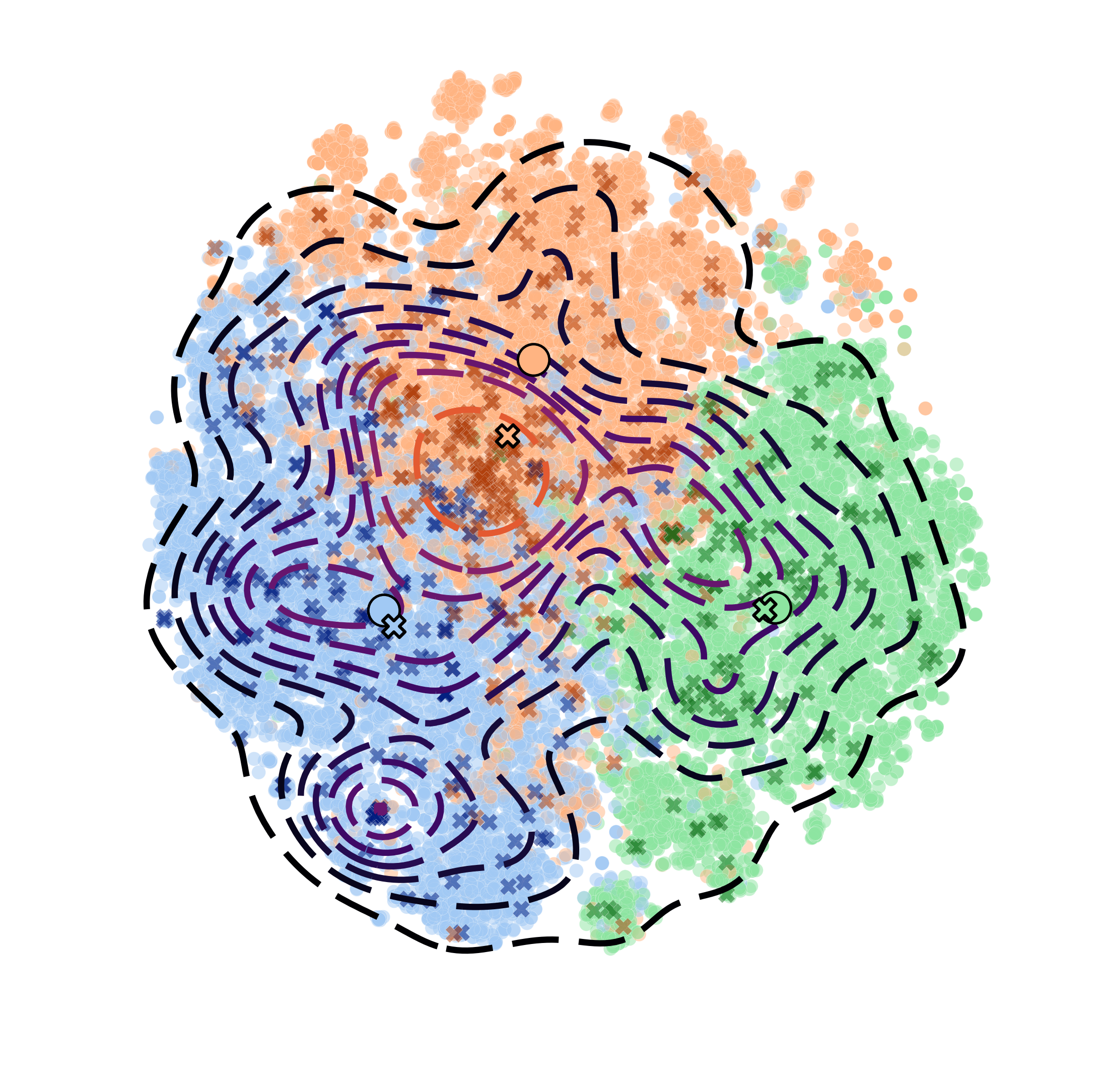}
        \caption{t-SNE \citep{t-SNE} visualisation for \kmeans clustering ($n=3$) of \mind dataset, discussed in~\refsec{cse-clustering}. Please refer to \refapp{appendix:num-clusters-cse} for plots of other datasets. 
        }
        \label{fig:mind-clustering-with-contour-plots}
\end{figure}

\paragraph{Clustering} 
\label{cse-clustering}
We first employ clustering algorithms to organize the embeddings in the dataset (which attackers have retrieved) into groups. This action enhances the subsequent selection step by: \textit{(i)} improving the efficiency of calculating pair-wise distance within smaller sets of embeddings, and \textit{(ii)} providing distinct groups of poisoned data entries, which facilitates the identification of more anomalous pairs. 
\kmeans algorithm~\citep{kmeans} is used as the primary clustering approach, while we discuss the effectiveness of other clustering methods in \refapp{appendix:clustering-algo-comparision}. Nevertheless, clustering solely is not sufficient for filtering out the watermarked embeddings. For instance, we can observe from the contour lines in~\reffig{fig:mind-clustering-with-contour-plots} that watermarked samples are spread across clusters and inconspicuous. Furthermore, the centroids of the watermarked samples and overall clusters do not coincide. To counteract this, we thus propose the selection module to identify the most suspicious embeddings with the watermark.

\begin{figure}[!htb]
    \centering    \includegraphics[width=0.75\columnwidth,keepaspectratio]{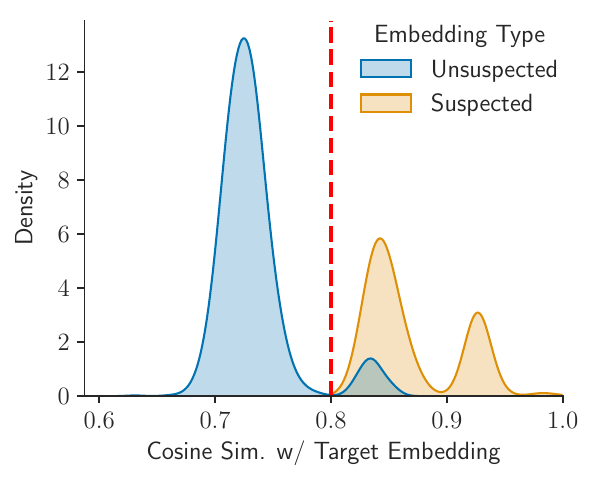}
    \caption{Similarity distribution plot between the target embedding and various embedding types. As we can see, the suspected embeddings returned by the selection module in \ourattack are distinctly different from unsuspected embeddings and more akin to the target embedding. The results for other datasets are reported in \refapp{appendix:other-cos-sim-dist-plots}. 
    }
    \label{fig:sst2-cos-sim-dist-sim-with-target-emb}
\end{figure}

\paragraph{Selection}
We denote the victim model as $\Theta_v$ and introduce another hold-out standard model (or benchmark model) as $\Theta_s$. %
Within each cluster $\mathcal C_i $, we conduct pairwise evaluations on the corresponding embeddings $\ve_{p}$ (provided embedding) and $\ve_{s}$ (standard embedding). Those with distinctive distance changes are considered suspected samples.

\prevWM incorporates varying proportions of a predetermined target embedding into texts containing trigger words. Since the predefined target embedding lacks shared semantic meaning, we hypothesize that the distance between embedding pairs, which have notable contributions from watermarks, exhibit anomaly behavior (validated in ~\reffig{fig:sst2-cos-sim-dist-sim-with-target-emb}) compared to corresponding distances derived from a standard language model, such as BERT \citep{devlin-etal-2019-bert}. Such difference is used to reflect the significance of the abnormal rank of watermarked pairs, as estimated by %
\begin{equation} 
\label{eq:percentile}
    \begin{aligned}
    D_p = \text{Rank}(D_v) - \text{Rank}(D_s), \\
    \end{aligned}
\end{equation}
where $D_v$ and $D_s$ are cosine similarity disparities between victim embeddings (\ie, $\ve_{p1}$ and $\ve_{p2}$) potentially containing watermark, and the standard embeddings \footnote{We use state-of-the-art SBERT \citep{SBERT} embeddings offering benefits in distance measurement.}
(\ie, $\ve_{s1}$ and $\ve_{s2}$) probably without watermarks. The $Rank$ function indicates the rank of the similarity scores by the embedding pairs in the set of scores ($D_v$ or $D_s$). The distinction between the suspected and unsuspected embeddings is illustrated in \reffig{fig:sst2-cos-sim-dist-sim-with-target-emb}.

\paragraph{Elimination}
Given the suspicious embeddings that are potentially watermarked from the previous step, we hypothesize that the watermark can be identified and recovered in suspicious embeddings' top principal components (validated in \refsec{sec:target-emb-reconstruction}) because the target embedding would be common among them. Following this idea, we propose an elimination algorithm, composed of two steps. First, we apply singular value decomposition (SVD) \cite{SVD} to analyze and identify the top $K$ principal components. Then, the contribution of these components are iteratively eliminated using the Gram-Schmidt (GS) process ~\cite{Gram-Schmidt}. The elimination step for each principal component vector is demonstrated as follows: 

\begin{equation}    
\vu^{\langle k+1\rangle} = \ve^{\langle k\rangle} - \text{Proj}(\ve^{\langle k\rangle}, \vc^{\langle k\rangle}).
\end{equation}

In the projection function~\cite{Projection}, $\ve^{\langle k\rangle}$ is projected onto the $k$-th principal component $\vc^{\langle k\rangle}$, denoted as 
\begin{equation}
\text{Proj}(\ve^{\langle k\rangle}, \vc^{\langle k\rangle}) = \frac{\vc^{\langle k\rangle} \cdot \ve^{\langle k\rangle}}{||\vc^{\langle k\rangle}||} \cdot \vc^{\langle k\rangle}.
\end{equation}
Here, $\ve^{\langle 0 \rangle}$ is initialized with $\ve_p$. After each iteration, $\ve^{\langle k+1\rangle}$ is acquired by normalizing $\vu^{\langle k+1\rangle}$ such that $ \text{Norm}\left(\vu^{\langle k+1\rangle} \right) = {\vu^{\langle k+1\rangle}}/{|| \vu^{\langle k+1\rangle} ||}$.

\subsection{\ourdefence Defense Framework} \label{multi-watermark}
In response to the successful \ourattack attack, we propose \textbf{W}atermark \textbf{A}ugmentation for \textbf{R}obust \textbf{DE}fe\textbf{N}se (\ourdefence) as a counter measurement, which incorporates multiple directions as watermarking embeddings. 

\paragraph{Multi-Directional Watermarks} 

To diversify the possibility of watermark directions (or target embeddings), we introduce multiple watermarks, noted as $\mW=\{\vw_{1}, \vw_{2}, ..., \vw_{R}\}$. This strategy increases the difficulty of inferring all of them via the elimination module in \ourattack attack. These watermarks remain confidential on servers and can be subject to regular updates. We randomly split the trigger words set, $T$ into $R$ independent subsets $T_r$ for $R$ watermarks. Then the trigger counting function, $\lambda_{r}$ is the frequency of trigger words in $T_{r}$ set with a maximum threshold of $m$ (level of watermark).
Finally, we add watermarks to the original embedding $\ve_o$ for text $S$ to generate the corresponding embedding $\ve_p$ as follows:

\begin{equation}
\label{eq:warden-watermark-augmentation}
\small
\begin{aligned}
    \text{Norm}\left( (1- \sum_{r=1}^{R} \lambda_r(S)) \cdot \ve_o+\sum_{r=1}^{R} \lambda_r(S) \cdot \vw_{r} \right).
\end{aligned}
\end{equation}
One thing to note is that because we split $T$ for multiple watermarks, the proportion of watermarked samples is independent of $R$ and is the same as in the single watermark case. Due to weight values being implicitly normalised $\lambda(S) = \lambda_1(S) + \lambda_2(S) + … + \lambda_R(S) $, where $\lambda(S)$ is watermark weight used on a single trigger set. 

\paragraph{Multi-Watermark Verification}
\label{sec:warden-copyright-verification}
We adopt a conservative approach to copyright verification with multiple watermarks, \ie, if any watermark confidently flags IP infringement, we consider it positive. Hence, we build verification datasets, backdoor texts $D_{b_r}$ and benign text $D_{n}$ as follows:
\begin{equation}
\centering
\begin{aligned}
    D_{b_r} &= \{ [ t_1, t_2, ..., t_m ] | t_i \in T_r \}, \forall r \in [1 .. R ], \\
    D_{n} &= \{ [ t_1, t_2, ..., t_m ] | t_i \notin T \}.
\end{aligned}
\end{equation}

The premise is that embeddings for these backdoor texts will be closer to their corresponding target embedding in contrast to benign texts in the case of watermarks. We leverage this behavior of embedding backdoors to verify copyright infringement at each watermark level. We quantify the closeness by computing cosine similarity and squared $L_2$ distance between target embeddings $\mW$ and embeddings of $D_{b_r}$ and $D_{n}$, \ie,  

\begin{equation}
\centering
\begin{aligned}
    \cos_{ir} &= \frac{\ve_i \cdot \vw_{r}}{|| \ve_i|| \cdot ||\vw_{r} ||}, \quad l_{2ir} = \biggl|\biggl| \frac{\ve_i}{|| \ve_i ||} - \frac{\vw_{r}}{|| \vw_{r} ||} \biggl|\biggl|^2, \\
    C_{b_r} &= \{\cos_{ir} | i \in D_{b_r} \},
    C_{n_r} = \{\cos_{ir} | i \in D_{n} \},  \\
    L_{b_r} &= \{l_{2ir} | i \in D_{b_r} \}, 
    L_{n_r} = \{l_{2ir} | i \in D_{n} \}, 
\end{aligned}
\end{equation}
where $r \in [1 .. R ]$.

The copyright detection performance is evaluated by taking the difference of averaged cosine similarity and averaged squared $L_2$ distance as per \refeq{delta-metrics-warden}. Furthermore, we compute the $\text{p-value}_j$ using the Kolmogorov-Smirnov (KS) test \cite{KStest} as the third metric, which compares these test value distributions. We aim to reject the null hypothesis: \textit{The two cosine similarity value sets $C_{b_r}$ and $C_{n_r}$ are consistent.}
\begin{equation}
\label{delta-metrics-warden}
\centering
\begin{aligned}
    \Delta_{\cos_r} &= \frac{1}{|C_{b_r}|}\sum_{i \in C_{b_r}} i - \frac{1}{|C_{n_r}|}\sum_{j \in C_{n_r}} j , \\
    \Delta_{l2_{r}} &= \frac{1}{|L_{b_r}|}\sum_{i \in L_{b_r}} i - \frac{1}{|L_{n_r}|}\sum_{j \in L_{n_r}} j . \\
\end{aligned}
\end{equation}

We evaluate these three metrics independently for all the watermarks and then combine them,
\begin{equation}
\begin{aligned}
    \Delta_{\cos} &= \max\limits_{1\leq r\leq R}{\Delta_{\cos_r}}, \\ 
    \Delta_{l2} &= \min\limits_{1\leq r\leq R}{\Delta_{l2_{r}}}, \\ 
    \text{p-value} &= \min\limits_{1\leq r\leq R}{\text{p-value}_r}.
\end{aligned}
\label{eq:warden-delta-calc}
\end{equation}
The core idea is that overall infringement can be certified by the infringement of any one of the target watermarking embeddings. 

\section{Experiments}

\subsection{Experimental Settings}
\paragraph{Evaluation Dataset}
To benchmark our attack and defense, we employ standard NLP datasets: \enron \cite{Enron-dataset}, \sst \cite{sst2-dataset}, \agnews \cite{ag_news-dataset}, and \mind \cite{MIND-dataset}. We use \enron dataset for email spam classification. \agnews and \mind are news-based and used for recommendation and classification tasks. We use \sst for sentiment classification. The statistics of these datasets are reported in \reftab{table:dataset-statistics}. 

\begin{table}[ht]
\centering
    \begin{tabular}{ccccc}
    
        \toprule
        {Dataset} & {\# Train} & {\# Test} & {\# Class}\\
        
        \midrule
        {\sst} & {67,349} & {872} & {2} \\
        {\mind} & {97,791} & {32,592} & {18} \\
        {\agnews} & {120,000} & {7,600} & {4} \\
        {\enron} & {31,716} & {2,000} & {2}\\
        \bottomrule
    \end{tabular}
    \caption{Statistics for classification datasets.}
    \label{table:dataset-statistics}
\end{table}

\paragraph{Evaluation Metrics}
To evaluate different aspects of our techniques, we adopt the following metrics:
\begin{itemize}
    \item \textbf{(Downstream) Task Performance} We construct a multi-layer perceptron (MLP) classifier with the EaaS embeddings as inputs. The quality of the embeddings is measured by the accuracy and \fscore of the classifiers on the downstream tasks. %
    \item \textbf{(Reconstruction) Attack Performance} We measure the closeness of reconstructed target embedding(s) (more details in \refsec{sec:target-emb-reconstruction}) with original target embedding(s) by reporting their cosine similarity.
     \item \textbf{(Infringement) Detection Performance} Following previous work \citep{peng-etal-2023-copying}, we employ three metrics, i.e., p-value, difference of cosine similarity, and difference of squared $L_2$ distance. Their customized variations for \ourdefence are defined in \refsec{sec:warden-copyright-verification}. Our findings largely rely on this evaluation as it reflects the performance in real-world applications.
\end{itemize}
\paragraph{Experimental environment} is detailed in the \refapp{appendix:exp-setting}.

\subsection{\ourattack Experiments}
\label{cse-attack-exps}
\ourattack is designed to assist model extraction attack bypassing post-publish copyright verification.
Hence, we evaluate whether we are able to bypass the copyright verification using the same watermark detection metrics with an opposite objective, \ie, lower p-value and the absolute values of $\Delta$ metrics close to zero. 

\paragraph{Watermark Elimination}
\begin{table}[t!]
    \centering
    \begin{tabular}{cccc}
        \toprule
        \multirow{2}{*}{Dataset} & \multicolumn{3}{c}{Detection Performance} \\
        \cmidrule(lr){2-4}
        {} & {p-value} & {$\Delta_{cos}(\%)$} & {$\Delta_{l2}(\%)$} \\
        \hline
        {\sst} & $> 0.83$ & 0.00 & 0.01 \\
        {\mind} & $> 0.57$ & 0.00 & 0.00 \\
        {\agnews} & $> 0.57$ & 0.09 & -0.18 \\
        {\enron} & $> 0.17$ & 0.00 & 0.01 \\
        \bottomrule
    \end{tabular}
    \caption{Copyright verification can be bypassed when the target direction is known and eliminated 
    from the provided embeddings.}
    \label{table:remove-target-direction}
\end{table}

One of the critical elements in the \prevWM is the secret target embedding ($\vw$) used for adding the watermark. The objective of \ourattack is to recover and erase this direction from the provided embeddings to circumvent copyright verification. We show later (see \reftab{table:attack-performance}) that such elimination is feasible and does not deteriorate the EaaS quality. To demonstrate that, we start with a simplified case where we assume access to this target embedding and directly remove this direction. Expectedly, as seen in \reftab{table:remove-target-direction}, we can bypass the copyright verification with minimal impact on the downstream utility performance. Moreover, this validates the importance of the projection technique employed in \ourattack. 
Additionally, this raises another technique's vulnerability of ensuring target embedding is kept secure. %

\paragraph{Watermark Reconstruction}
\label{sec:target-emb-reconstruction}
In a successful attack, the principal components $\vc^{\langle k \rangle}$ removed from the embeddings erase the watermark by recovering the target embedding. To validate this conjecture, we model and solve an optimization problem as defined in \refeq{eq:reconstruction-target-emb} where a linear combination of $\vc^{\langle k \rangle}$ results in the recovered target embedding $\vw$. We then calculate cosine similarity to the target embedding $\vw$. A high cosine similarity demonstrates the \ourattack technique's effectiveness. For \ourattack, the reconstructed target embedding is extremely (99+\% cosine similarity) close to the original target embedding (more in following \refsec{sec:cse-results}), 
\begin{align}
\label{eq:reconstruction-target-emb}
    \min_{\pmb{\alpha}} \biggl\|\vw - \sum_{k=1}^{K}{\alpha}_k \cdot \vc^{\langle k\rangle}\biggl\|^2.
\end{align}

\paragraph{Effectiveness Evaluation}
\label{sec:cse-results}
\begin{table*}
\centering
    \begin{minipage}{1.\textwidth}
    \resizebox{\textwidth}{!}{%
    \begin{tabular}{ccccccc}
    \toprule
    \multirow{2}{*}{Dataset} & \multirow{2}{*}{Method} & \multicolumn{2}{c}{Task Performance} & \multicolumn{3}{c}{Detection Performance} \\
    \cmidrule(lr){3-4} \cmidrule(lr){5-7}
    {} & {} & ACC.(\%) & \fscore & {p-value $\uparrow$} & {$\Delta_{cos}(\%) \downarrow$} & {$\Delta_{l2}(\%) \uparrow$}\\
    \midrule
    \multirow{3}{*}{\sst} & Original & 93.42$\pm$0.13 & 93.42$\pm$0.13 & $>0.47$ & -0.18$\pm$0.22 & 0.37$\pm$0.43 \\
    & \prevWM & 93.12$\pm$0.12 & 93.12$\pm$0.12 & {$< 10^{-3}$} & 3.56$\pm$0.50 & -7.11$\pm$1.01 \\
    
     & \prevWM + \ourattack & 90.46$\pm$0.98 & 90.46$\pm$0.98 & $> 0.04$ & \textbf{0.99}$\pm$0.40 & \textbf{-1.97}$\pm$0.80 \\
    \midrule
    \multirow{3}{*}{\mind} & Original & 77.22$\pm$0.13 & 51.37$\pm$0.31 & $>0.26$ & -0.69$\pm$0.17 & 1.37$\pm$0.35 \\
    &\prevWM & 77.19$\pm$0.09 & 51.40$\pm$0.16 & {$< 10^{-6}$} & 4.69$\pm$0.17 & -9.37$\pm$0.33 \\
    
     & \prevWM + \ourattack& 75.51$\pm$0.16 & 50.35$\pm$0.46 & $> 0.21$ & \textbf{0.55}$\pm$0.18 & \textbf{-1.10}$\pm$0.37 \\
    \midrule
    \multirow{3}{*}{\agnews} & Original& 93.64$\pm$0.11 & 93.64$\pm$0.11 & $>0.36$ & 0.56$\pm$0.24 & -1.13$\pm$0.48 \\
    
     &\prevWM & 93.52$\pm$0.11 & 93.52$\pm$ 0.11 & {$< 10^{-9}$} & 12.76$\pm$0.43 & -25.52$\pm$0.87 \\
    
     & \prevWM + \ourattack& 92.87$\pm$0.32 & 92.87$\pm$0.32 & $> 0.22$ & \textbf{0.27}$\pm$0.30 & \textbf{-0.55}$\pm$0.60 \\
    \midrule
    \multirow{3}{*}{\enron} & Original& 94.73$\pm$0.14 & 94.73$\pm$0.14 & $>0.20$ & -0.38$\pm$0.38 & 0.76$\pm$0.75 \\
    &\prevWM & 94.61$\pm$0.28 & 94.61$\pm$0.28 & {$< 10^{-6}$} & 5.93$\pm$0.28 & -11.86$\pm$0.56 \\
     & \prevWM + \ourattack& 95.56$\pm$0.21 & 95.56$\pm$0.21 & $> 0.62$ & \textbf{0.59}$\pm$0.33 & \textbf{     
     -1.17}$\pm$0.65 \\
    
    \bottomrule
    \end{tabular}}
    \caption{The performance of \ourattack for different scenarios on \sst, \mind, \agnews, and \enron datasets. `Original' represents a benign victim model, `\prevWM' stands for the existing watermarking technique, and `$\text{\prevWM}+\text{\ourattack}$' is the case where \ourattack is performed on provided embeddings by \prevWM before doing model extraction (as shown in ~Figure~\ref{fig:warden-overview}.c). $\uparrow$ denotes higher metrics are better and $\downarrow$ denotes lower metrics are better from the attacker's objective. } 
    \label{table:attack-performance}
    \end{minipage}
\end{table*}

In \reftab{table:attack-performance}, \ourattack along with model extraction attack is proved effective in removing the influence from \prevWM. Detection performance dropped to almost the original case, which indicates the EaaS provider will not be able to detect the imitation attack performed by the attacker. In addition, for \sst, \mind, \agnews datasets, the downstream performance dropped 1-2\%, which demonstrates the attack preserves the embeddings utility. 
We skip attack performance in the \reftab{table:attack-performance} as they all have (almost) full watermark reconstruction. However, we discuss this in \refsec{sec:warden-exps} where we observe varying values due to the ineffectiveness of \ourattack attack against \ourdefence defense.

\paragraph{Ablation Study}
\begin{table}
    \begin{minipage}{0.95\columnwidth}
    \resizebox{\textwidth}{!}{%
    \begin{tabular}{cccccc}
    \toprule
    \multirow{2}{*}{Dataset} & \multicolumn{2}{c}{Task Performance} & \multicolumn{3}{c}{Detection Performance} \\
    \cmidrule(lr){2-3} \cmidrule(lr){4-6}
    {} & ACC.(\%) & \fscore & p-value & {$\Delta_{cos}(\%)$} & {$\Delta_{l2}(\%) $}\\
    \toprule
    
    \sst & 87.04 & 87.01 & $> 0.05$ & 0.19 & -0.39 \\
    \mind & 74.80 & 50.57 & $> 0.08$ & 1.09 & -2.19 \\
    \agnews & 93.04 & 93.04 & $> 0.01$ & -2.14 & 4.29 \\
    \enron & 95.45 & 95.45 & $> 0.17$ & -1.28 & 2.57 \\

    \bottomrule
    \end{tabular}}
    \caption{\ourattack on a non-watermarked victim model, with minimal degradation in downstream utility and copyright detection metrics of an innocent model.}
    \label{table:attack-on-non-watermarked-model}
    \end{minipage}
\end{table}

An attacker will not be aware whether the model they are trying to imitate is watermarked. \reftab{table:attack-on-non-watermarked-model} shows that our attack leads to only minor quality degradation for such scenarios, demonstrating the suitability of \ourattack. We perform further extensive quantitative and qualitative sensitivity study to investigate how other factors (such as algorithms, parameters, and models) affect the efficacy of our suggested \ourattack attack in \refapp{appendix:cse-ablation}.

\subsection{\ourdefence Experiments}
\label{sec:warden-exps}
\paragraph{Watermarking Performance of \ourdefence}
\label{sec:warden-perf}

\begin{figure}[h]
    \centering    
    \includegraphics[width=0.95\columnwidth,keepaspectratio]{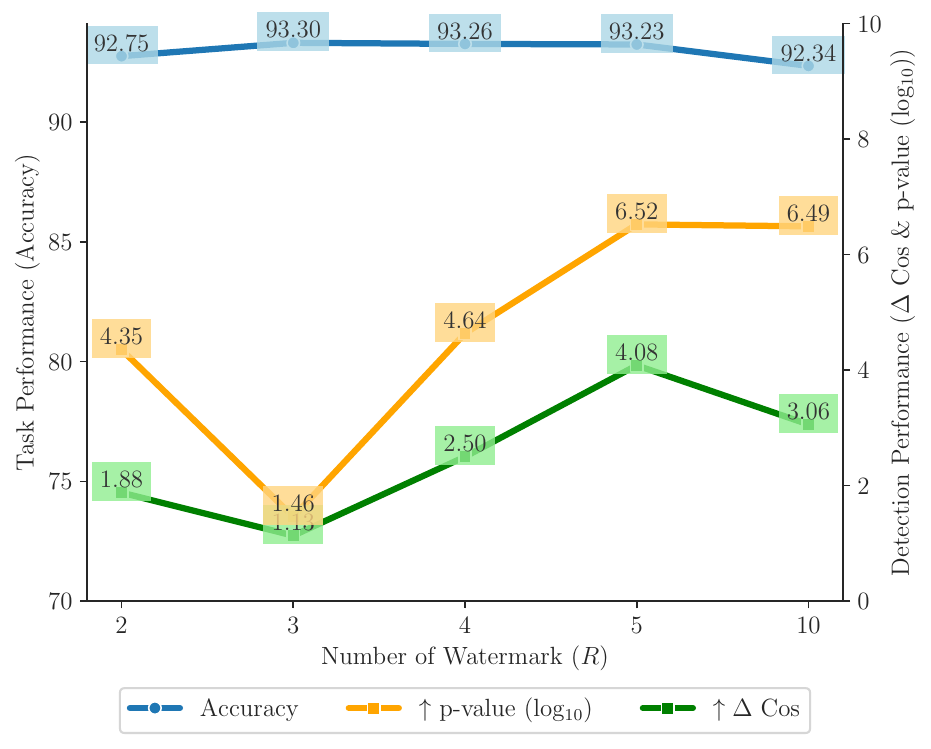}
    \caption{The impact of the number of watermarks ($R$) in \ourdefence for \sst dataset. 
    }
    \label{fig:warden-sst2}
\end{figure}

We illustrate the efficiency of employing multiple watermarks in~\reffig{fig:warden-sst2}, which demonstrates the outstanding performance (yellow and green line upward trend) of \ourdefence with increasing $R$ and marginal degradation (blue line) in the downstream utility. The results on other datasets also show similar patterns which can be found in \refapp{warden-other-dataset-results}.
\begin{figure}[t]
    \centering    \includegraphics[width=0.95\columnwidth,keepaspectratio]{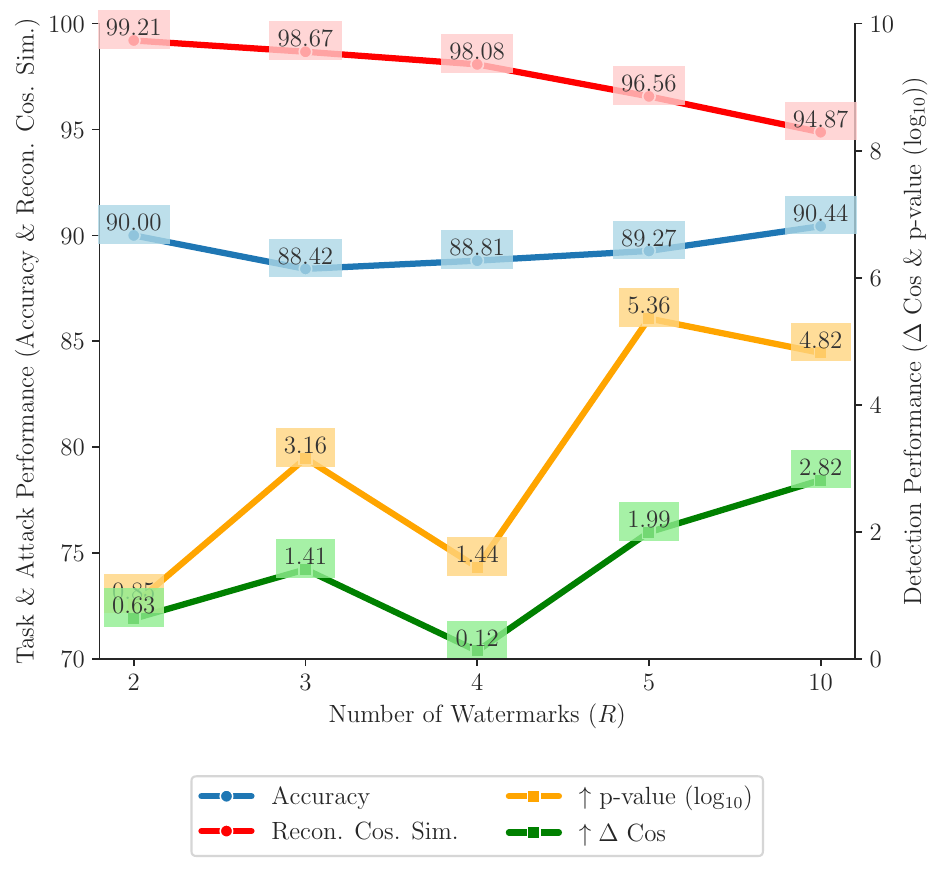}
    \caption{The impact of the number of watermarks ($R$) in \ourdefence against \ourattack on \sst dataset. 
    Note: `Recon. Cos. Sim.' (the \textcolor{red}{Red} line) represents the minimum reconstructed cosine similarity among all possible watermarks in $\mW$.
    }
    \label{fig:CSE-on-WARDEN-sst2}
    \vspace{-5mm}
\end{figure}

\paragraph{\ourdefence against \ourattack}
\label{sec:attack-on-our-defence}

Now, we investigate the effectiveness of \ourdefence against \ourattack (shown in \reffig{fig:CSE-on-WARDEN-sst2}). As observed in the previous section, \ourdefence is stealthier with increasing watermarks. As expected, the performance of \ourattack diminishes, correlating with decreasing attack performance (red line). Due to the usage of more watermarks, there is a natural increase in the likelihood that one of them will detect an infringement. Moreover, in extreme scenarios, a mixture of multiple target embeddings will substitute the watermarked samples (\refeq{eq:warden-watermark-augmentation}), reducing the impact of the \ourattack attack's exploitation of the semantic distortion in the embeddings. 

\paragraph{Gram-Schmidt Extension} 
\begin{figure}[!htbp]
    \centering    \includegraphics[width=0.95\columnwidth,keepaspectratio]{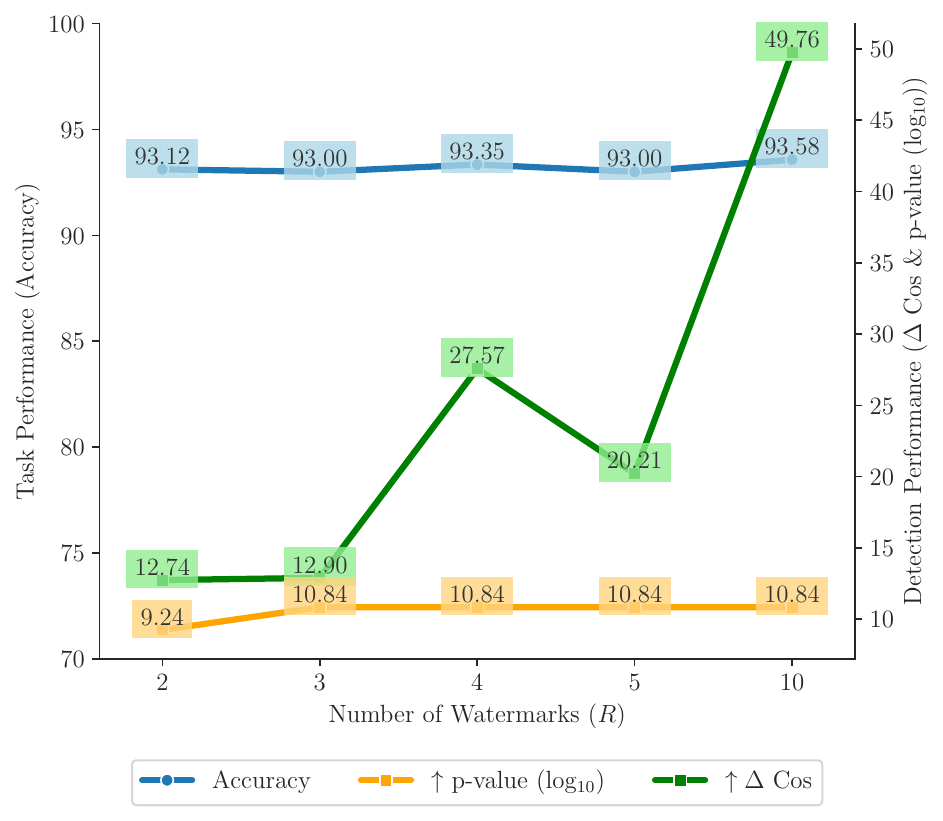}
    \caption{The impact of GS extension on \ourdefence for \sst dataset. The observations are in line with normal \ourdefence (\reffig{fig:CSE-on-WARDEN-sst2}) results with the only difference being stronger metrics.
    }
    \label{fig:GSO-defence-sst2}
\end{figure}

To further strengthen \ourdefence, we investigate the application of the Gram-Schmidt (GS) process \citep{Gram-Schmidt} on target embeddings $\mW$, as we assume the orthogonal set of watermark embeddings are more distinguishable to each other. In our experiments, as reported in ~\reffig{fig:GSO-defence-sst2}, the detection performance is stronger after GS selection. In addition, due to orthogonality, the reconstructed target embedding cosine similarities will be significantly lower, indicating \ourattack might also be ineffective. We observe the same from the corresponding ablation study in \refapp{warden-other-dataset-results}.

\paragraph{Ablation Study}

Similar to the experiments for \ourattack, we perform \ourdefence on non-watermarked models. Due to our strict verification, for the high value of $R$, the p-value could be noisy. It is because the verification process might find closeness due to genuine semantics instead of backdoors as a result of a high pool of watermark directions. This could lead to false positives, \ie, incorrectly classifying models as copied. However, in such cases, we observe that other detection metrics ($\Delta$ based) metrics are reliable, which should aid the entity in making appropriate decisions (refer \reffig{fig:warden-non-watermarked}). We conduct a further detailed ablation study dissecting the \ourdefence components and showing its stealthiness in \refapp{appendix:ablation-our-defence}.

\section{Conclusion}
In this paper, we first demonstrate that our new \ourattack attack can bypass the recent EaaS watermarking technique. \ourattack cleanses the watermarked dataset by clustering them first, then selecting embedding pairs with disparity, and finally eliminating their top principal components, while maintaining the service utility. To remedy this shortcoming, we propose a simple yet effective watermarking method, \ourdefence, which augments the previous approach by introducing multiple watermarks to embeddings. Our intensive experiments show that \ourdefence is superior in verifying the copyright of EaaS from prior works. Furthermore, \ourdefence is also effective against potent \ourattack, which shows its resilience to different attacks. We also conduct detailed ablation studies to verify the importance of every component of \ourattack and \ourdefence. Future studies may consider exploring watermark ownership under multi-owner service settings.

\section*{Limitations}
We test our \ourdefence defense against \ourattack attack, acknowledging that various other attacks might overcome the uni-directional watermark approach.
Although publishing the \ourdefence algorithm to the public may inspire future attacks against it, we do not foresee it to be a trivial task, as the capability of \ourdefence can be enhanced by using more conservative strategies, \eg, more stealthy trigger patterns and watermarking techniques.

We also know that by having access to the ground-truth watermarking vectors, combined with the GS process, one can eliminate the \ourdefence watermarks, as discussed in \refapp{appendix:warden-remove-target-embs}. However, it is the service provider's responsibility to maintain the confidentiality of their watermarking keys. 

We also note the false positive of the p-value for non-watermarked models when a large number of watermark vectors are augmented (\reffig{fig:warden-non-watermarked}), even though other metrics rectify this incorrect signal. We suggest service providers conduct a preliminary study to select the number of watermarks for \ourdefence. In this regard, the current work is an empirical observation study, and theoretical analysis might help decide the optimal number of watermarks. In future, we will endeavour to investigate advanced watermarking mechanisms focusing on defence purposes.

\section*{Social Impact}
We developed a \ourattack attack, which could aid attackers in circumventing EaaS IP infringements. We agree that with \ourattack, any existing system using \prevWM is vulnerable. Yet, we argue that it is critical to show the possibility of such attacks and make users aware of them. The usual first step in security is to first expose the vulnerability. Additionally, to mitigate the aforementioned threat, we contribute an improved watermarking technique, \ourdefence, which could be incorporated with minimal effort.

\section*{Acknowledgements}
We would like to appreciate the valuable feedback from all anonymous reviewers. This research was supported by The University of Melbourne’s Research Computing Services and the Petascale Campus Initiative. Qiongkai Xu would like to express his gratitude to the FSE Staff Travel Scheme and FSE DDRI grant for the support in both travel and research. Anudeex Shetty would like to express his gratitude to Newman College for the support in travel and research.

\bibliography{custom}

\begin{thebibliography}{47}
\expandafter\ifx\csname natexlab\endcsname\relax\def\natexlab#1{#1}\fi

\bibitem[{Alzantot et~al.(2018)Alzantot, Sharma, Elgohary, Ho, Srivastava, and Chang}]{alzantot-etal-2018-generating}
Moustafa Alzantot, Yash Sharma, Ahmed Elgohary, Bo-Jhang Ho, Mani Srivastava, and Kai-Wei Chang. 2018.
\newblock \href {https://doi.org/10.18653/v1/D18-1316} {Generating natural language adversarial examples}.
\newblock In \emph{Proceedings of the 2018 Conference on Empirical Methods in Natural Language Processing}, pages 2890--2896, Brussels, Belgium. Association for Computational Linguistics.

\bibitem[{Arthur et~al.(2007)Arthur, Vassilvitskii et~al.}]{kmeans}
David Arthur, Sergei Vassilvitskii, et~al. 2007.
\newblock k-means++: The advantages of careful seeding.
\newblock In \emph{Soda}, volume~7, pages 1027--1035.

\bibitem[{Berger and Zhou(2014)}]{KStest}
Vance~W. Berger and YanYan Zhou. 2014.
\newblock \href {https://doi.org/https://doi.org/10.1002/9781118445112.stat06558} {\emph{Kolmogorov–Smirnov Test: Overview}}. John Wiley \& Sons, Ltd.

\bibitem[{Brown et~al.(2020)Brown, Mann, Ryder, Subbiah, Kaplan, Dhariwal, Neelakantan, Shyam, Sastry, Askell, Agarwal, Herbert-Voss, Krueger, Henighan, Child, Ramesh, Ziegler, Wu, Winter, Hesse, Chen, Sigler, Litwin, Gray, Chess, Clark, Berner, McCandlish, Radford, Sutskever, and Amodei}]{GPT}
Tom Brown, Benjamin Mann, Nick Ryder, Melanie Subbiah, Jared~D Kaplan, Prafulla Dhariwal, Arvind Neelakantan, Pranav Shyam, Girish Sastry, Amanda Askell, Sandhini Agarwal, Ariel Herbert-Voss, Gretchen Krueger, Tom Henighan, Rewon Child, Aditya Ramesh, Daniel Ziegler, Jeffrey Wu, Clemens Winter, Chris Hesse, Mark Chen, Eric Sigler, Mateusz Litwin, Scott Gray, Benjamin Chess, Jack Clark, Christopher Berner, Sam McCandlish, Alec Radford, Ilya Sutskever, and Dario Amodei. 2020.
\newblock \href {https://proceedings.neurips.cc/paper_files/paper/2020/file/1457c0d6bfcb4967418bfb8ac142f64a-Paper.pdf} {Language models are few-shot learners}.
\newblock In \emph{Advances in Neural Information Processing Systems}, volume~33, pages 1877--1901. Curran Associates, Inc.

\bibitem[{Chandrasekaran et~al.(2020)Chandrasekaran, Chaudhuri, Giacomelli, Jha, and Yan}]{steal-from-API-1}
Varun Chandrasekaran, Kamalika Chaudhuri, Irene Giacomelli, Somesh Jha, and Songbai Yan. 2020.
\newblock Exploring connections between active learning and model extraction.
\newblock In \emph{Proceedings of the 29th USENIX Conference on Security Symposium}, SEC'20, USA. USENIX Association.

\bibitem[{Chen et~al.(2022)Chen, Meng, Sun, Guo, Zhang, Li, and Fan}]{chen2022badpre}
Kangjie Chen, Yuxian Meng, Xiaofei Sun, Shangwei Guo, Tianwei Zhang, Jiwei Li, and Chun Fan. 2022.
\newblock \href {https://openreview.net/forum?id=Mng8CQ9eBW} {Badpre: Task-agnostic backdoor attacks to pre-trained {NLP} foundation models}.
\newblock In \emph{International Conference on Learning Representations}.

\bibitem[{Dai et~al.(2019)Dai, Chen, and Li}]{backdoor-attack}
Jiazhu Dai, Chuanshuai Chen, and Yufeng Li. 2019.
\newblock \href {https://doi.org/10.1109/ACCESS.2019.2941376} {A backdoor attack against lstm-based text classification systems}.
\newblock \emph{IEEE Access}, 7:138872--138878.

\bibitem[{Devlin et~al.(2019)Devlin, Chang, Lee, and Toutanova}]{devlin-etal-2019-bert}
Jacob Devlin, Ming-Wei Chang, Kenton Lee, and Kristina Toutanova. 2019.
\newblock \href {https://doi.org/10.18653/v1/N19-1423} {{BERT}: Pre-training of deep bidirectional transformers for language understanding}.
\newblock In \emph{Proceedings of the 2019 Conference of the North {A}merican Chapter of the Association for Computational Linguistics: Human Language Technologies, Volume 1 (Long and Short Papers)}, pages 4171--4186, Minneapolis, Minnesota. Association for Computational Linguistics.

\bibitem[{Ebrahimi et~al.(2018)Ebrahimi, Rao, Lowd, and Dou}]{ebrahimi-etal-2018-hotflip}
Javid Ebrahimi, Anyi Rao, Daniel Lowd, and Dejing Dou. 2018.
\newblock \href {https://doi.org/10.18653/v1/P18-2006} {{H}ot{F}lip: White-box adversarial examples for text classification}.
\newblock In \emph{Proceedings of the 56th Annual Meeting of the Association for Computational Linguistics (Volume 2: Short Papers)}, pages 31--36, Melbourne, Australia. Association for Computational Linguistics.

\bibitem[{Golub and Reinsch(1970)}]{SVD}
G.~H. Golub and C.~Reinsch. 1970.
\newblock \href {https://doi.org/10.1007/BF02163027} {Singular value decomposition and least squares solutions}.
\newblock \emph{Numer. Math.}, 14(5):403–420.

\bibitem[{He et~al.(2022{\natexlab{a}})He, Lyu, Chen, and Xu}]{he-etal-2022-extracted}
Xuanli He, Lingjuan Lyu, Chen Chen, and Qiongkai Xu. 2022{\natexlab{a}}.
\newblock \href {https://doi.org/10.18653/v1/2022.emnlp-main.99} {Extracted {BERT} model leaks more information than you think!}
\newblock In \emph{Proceedings of the 2022 Conference on Empirical Methods in Natural Language Processing}, pages 1530--1537, Abu Dhabi, United Arab Emirates. Association for Computational Linguistics.

\bibitem[{He et~al.(2021{\natexlab{a}})He, Lyu, Sun, and Xu}]{he-etal-2021-model}
Xuanli He, Lingjuan Lyu, Lichao Sun, and Qiongkai Xu. 2021{\natexlab{a}}.
\newblock \href {https://doi.org/10.18653/v1/2021.naacl-main.161} {Model extraction and adversarial transferability, your {BERT} is vulnerable!}
\newblock In \emph{Proceedings of the 2021 Conference of the North American Chapter of the Association for Computational Linguistics: Human Language Technologies}, pages 2006--2012, Online. Association for Computational Linguistics.

\bibitem[{He et~al.(2021{\natexlab{b}})He, Lyu, Sun, and Xu}]{he-etal-2021-extraction-adversarial-transfer}
Xuanli He, Lingjuan Lyu, Lichao Sun, and Qiongkai Xu. 2021{\natexlab{b}}.
\newblock \href {https://doi.org/10.18653/v1/2021.naacl-main.161} {Model extraction and adversarial transferability, your {BERT} is vulnerable!}
\newblock In \emph{Proceedings of the 2021 Conference of the North American Chapter of the Association for Computational Linguistics: Human Language Technologies}, pages 2006--2012, Online. Association for Computational Linguistics.

\bibitem[{He et~al.(2022{\natexlab{b}})He, Xu, Lyu, Wu, and Wang}]{He_Xu_Lyu_Wu_Wang_2022}
Xuanli He, Qiongkai Xu, Lingjuan Lyu, Fangzhao Wu, and Chenguang Wang. 2022{\natexlab{b}}.
\newblock \href {https://doi.org/10.1609/aaai.v36i10.21321} {Protecting intellectual property of language generation apis with lexical watermark}.
\newblock \emph{Proceedings of the AAAI Conference on Artificial Intelligence}, 36(10):10758--10766.

\bibitem[{He et~al.(2022{\natexlab{c}})He, Xu, Zeng, Lyu, Wu, Li, and Jia}]{he2022cater}
Xuanli He, Qiongkai Xu, Yi~Zeng, Lingjuan Lyu, Fangzhao Wu, Jiwei Li, and Ruoxi Jia. 2022{\natexlab{c}}.
\newblock \href {https://openreview.net/forum?id=L7P3IvsoUXY} {{CATER}: Intellectual property protection on text generation {API}s via conditional watermarks}.
\newblock In \emph{Advances in Neural Information Processing Systems}.

\bibitem[{Huang et~al.(2023)Huang, Zhuo, Xu, Hu, Yuan, and Chen}]{huang2023training}
Yujin Huang, Terry~Yue Zhuo, Qiongkai Xu, Han Hu, Xingliang Yuan, and Chunyang Chen. 2023.
\newblock Training-free lexical backdoor attacks on language models.
\newblock In \emph{Proceedings of the ACM Web Conference 2023}, pages 2198--2208.

\bibitem[{Juuti et~al.(2019)Juuti, Szyller, Marchal, and Asokan}]{juuti2019prada}
Mika Juuti, Sebastian Szyller, Samuel Marchal, and N~Asokan. 2019.
\newblock Prada: protecting against dnn model stealing attacks.
\newblock In \emph{2019 IEEE European Symposium on Security and Privacy (EuroS\&P)}, pages 512--527. IEEE.

\bibitem[{Kirchenbauer et~al.(2023)Kirchenbauer, Geiping, Wen, Katz, Miers, and Goldstein}]{kirchenbauer2023watermark}
John Kirchenbauer, Jonas Geiping, Yuxin Wen, Jonathan Katz, Ian Miers, and Tom Goldstein. 2023.
\newblock A watermark for large language models.
\newblock In \emph{International Conference on Machine Learning}, pages 17061--17084. PMLR.

\bibitem[{Krishna et~al.(2020)Krishna, Tomar, Parikh, Papernot, and Iyyer}]{ModelExtraction_1}
Kalpesh Krishna, Gaurav~Singh Tomar, Ankur~P. Parikh, Nicolas Papernot, and Mohit Iyyer. 2020.
\newblock \href {https://openreview.net/forum?id=Byl5NREFDr} {Thieves on sesame street! model extraction of bert-based apis}.
\newblock In \emph{8th International Conference on Learning Representations, {ICLR} 2020, Addis Ababa, Ethiopia, April 26-30, 2020}. OpenReview.net.

\bibitem[{Li et~al.(2022)Li, Bai, Jiang, Yang, Xia, and Li}]{li2022untargeted}
Yiming Li, Yang Bai, Yong Jiang, Yong Yang, Shu-Tao Xia, and Bo~Li. 2022.
\newblock \href {https://openreview.net/forum?id=kcQiIrvA_nz} {Untargeted backdoor watermark: Towards harmless and stealthy dataset copyright protection}.
\newblock In \emph{Advances in Neural Information Processing Systems}.

\bibitem[{Lim et~al.(2022)Lim, Chan, Ng, Fan, and Yang}]{LIM2022108285}
Jian~Han Lim, Chee~Seng Chan, Kam~Woh Ng, Lixin Fan, and Qiang Yang. 2022.
\newblock \href {https://doi.org/https://doi.org/10.1016/j.patcog.2021.108285} {Protect, show, attend and tell: Empowering image captioning models with ownership protection}.
\newblock \emph{Pattern Recognition}, 122:108285.

\bibitem[{Liu et~al.(2022)Liu, Jia, Liu, and Gong}]{Eaas-Extraction-Attack}
Yupei Liu, Jinyuan Jia, Hongbin Liu, and {Neil Zhenqiang} Gong. 2022.
\newblock \href {https://doi.org/10.1145/3548606.3560586} {Stolenencoder: Stealing pre-trained encoders in self-supervised learning}.
\newblock In \emph{CCS 2022 - Proceedings of the 2022 ACM SIGSAC Conference on Computer and Communications Security}, Proceedings of the ACM Conference on Computer and Communications Security, pages 2115--2128. Association for Computing Machinery.

\bibitem[{Loshchilov and Hutter(2019)}]{loshchilov2018decoupled}
Ilya Loshchilov and Frank Hutter. 2019.
\newblock \href {https://openreview.net/forum?id=Bkg6RiCqY7} {Decoupled weight decay regularization}.
\newblock In \emph{International Conference on Learning Representations}.

\bibitem[{Merity et~al.(2016)Merity, Xiong, Bradbury, and Socher}]{WikiText-dataset}
Stephen Merity, Caiming Xiong, James Bradbury, and Richard Socher. 2016.
\newblock Pointer sentinel mixture models.
\newblock In \emph{ICLR}.

\bibitem[{Metsis et~al.(2006)Metsis, Androutsopoulos, and Paliouras}]{Enron-dataset}
Vangelis Metsis, Ion Androutsopoulos, and Georgios Paliouras. 2006.
\newblock Spam filtering with naive bayes-which naive bayes?
\newblock In \emph{CEAS}, volume~17, pages 28--69. Mountain View, CA.

\bibitem[{OpenAI(2024)}]{recentopenaiEmbeddingModels}
OpenAI. 2024.
\newblock {N}ew embedding models and {A}{P}{I} updates --- openai.com.
\newblock \url{https://openai.com/blog/new-embedding-models-and-api-updates}.
\newblock [Accessed 02-02-2024].

\bibitem[{Orekondy et~al.(2019)Orekondy, Schiele, and Fritz}]{ModelExtraction_2}
Tribhuvanesh Orekondy, Bernt Schiele, and Mario Fritz. 2019.
\newblock \href {https://doi.org/10.1109/CVPR.2019.00509} {Knockoff nets: Stealing functionality of black-box models}.
\newblock In \emph{2019 IEEE/CVF Conference on Computer Vision and Pattern Recognition (CVPR)}, pages 4949--4958.

\bibitem[{Pedregosa et~al.(2011)Pedregosa, Varoquaux, Gramfort, Michel, Thirion, Grisel, Blondel, Prettenhofer, Weiss, Dubourg, Vanderplas, Passos, Cournapeau, Brucher, Perrot, and Duchesnay}]{scikit-learn}
F.~Pedregosa, G.~Varoquaux, A.~Gramfort, V.~Michel, B.~Thirion, O.~Grisel, M.~Blondel, P.~Prettenhofer, R.~Weiss, V.~Dubourg, J.~Vanderplas, A.~Passos, D.~Cournapeau, M.~Brucher, M.~Perrot, and E.~Duchesnay. 2011.
\newblock Scikit-learn: Machine learning in {P}ython.
\newblock \emph{Journal of Machine Learning Research}, 12:2825--2830.

\bibitem[{Peng et~al.(2023)Peng, Yi, Wu, Wu, Bin~Zhu, Lyu, Jiao, Xu, Sun, and Xie}]{peng-etal-2023-copying}
Wenjun Peng, Jingwei Yi, Fangzhao Wu, Shangxi Wu, Bin Bin~Zhu, Lingjuan Lyu, Binxing Jiao, Tong Xu, Guangzhong Sun, and Xing Xie. 2023.
\newblock \href {https://doi.org/10.18653/v1/2023.acl-long.423} {Are you copying my model? protecting the copyright of large language models for {E}aa{S} via backdoor watermark}.
\newblock In \emph{Proceedings of the 61st Annual Meeting of the Association for Computational Linguistics (Volume 1: Long Papers)}, pages 7653--7668, Toronto, Canada. Association for Computational Linguistics.

\bibitem[{Perwass et~al.(2009)Perwass, Edelsbrunner, Kobbelt, and Polthier}]{Projection}
Christian Perwass, Herbert Edelsbrunner, Leif Kobbelt, and Konrad Polthier. 2009.
\newblock \emph{Geometric algebra with applications in engineering}, volume~4.
\newblock Springer.

\bibitem[{Radford et~al.(2019)Radford, Wu, Child, Luan, Amodei, Sutskever et~al.}]{radford2019language}
Alec Radford, Jeffrey Wu, Rewon Child, David Luan, Dario Amodei, Ilya Sutskever, et~al. 2019.
\newblock Language models are unsupervised multitask learners.
\newblock \emph{OpenAI blog}, 1(8):9.

\bibitem[{Reimers and Gurevych(2019)}]{SBERT}
Nils Reimers and Iryna Gurevych. 2019.
\newblock \href {https://doi.org/10.18653/v1/D19-1410} {Sentence-{BERT}: Sentence embeddings using {S}iamese {BERT}-networks}.
\newblock In \emph{Proceedings of the 2019 Conference on Empirical Methods in Natural Language Processing and the 9th International Joint Conference on Natural Language Processing (EMNLP-IJCNLP)}, pages 3982--3992, Hong Kong, China. Association for Computational Linguistics.

\bibitem[{Reynolds et~al.(2009)}]{GMM}
Douglas~A Reynolds et~al. 2009.
\newblock Gaussian mixture models.
\newblock \emph{Encyclopedia of biometrics}, 741(659-663).

\bibitem[{Socher et~al.(2013)Socher, Perelygin, Wu, Chuang, Manning, Ng, and Potts}]{sst2-dataset}
Richard Socher, Alex Perelygin, Jean Wu, Jason Chuang, Christopher~D. Manning, Andrew Ng, and Christopher Potts. 2013.
\newblock \href {https://aclanthology.org/D13-1170} {Recursive deep models for semantic compositionality over a sentiment treebank}.
\newblock In \emph{Proceedings of the 2013 Conference on Empirical Methods in Natural Language Processing}, pages 1631--1642, Seattle, Washington, USA. Association for Computational Linguistics.

\bibitem[{Tang et~al.(2023)Tang, Yu, Gai, Qu, Hu, Xiong, and Wu}]{VLPMarker}
Yuanmin Tang, Jing Yu, Keke Gai, Xiangyan Qu, Yue Hu, Gang Xiong, and Qi~Wu. 2023.
\newblock \href {http://arxiv.org/abs/2311.05863} {Watermarking vision-language pre-trained models for multi-modal embedding as a service}.

\bibitem[{Tram\`{e}r et~al.(2016)Tram\`{e}r, Zhang, Juels, Reiter, and Ristenpart}]{steal-from-API-2}
Florian Tram\`{e}r, Fan Zhang, Ari Juels, Michael~K. Reiter, and Thomas Ristenpart. 2016.
\newblock Stealing machine learning models via prediction apis.
\newblock In \emph{Proceedings of the 25th USENIX Conference on Security Symposium}, SEC'16, page 601–618, USA. USENIX Association.

\bibitem[{Trefethen and Bau(1997)}]{Gram-Schmidt}
Lloyd~N. Trefethen and David Bau. 1997.
\newblock \emph{Numerical Linear Algebra}.
\newblock SIAM.

\bibitem[{Uchida et~al.(2017)Uchida, Nagai, Sakazawa, and Satoh}]{Uchida}
Yusuke Uchida, Yuki Nagai, Shigeyuki Sakazawa, and Shin'ichi Satoh. 2017.
\newblock \href {https://doi.org/10.1145/3078971.3078974} {Embedding watermarks into deep neural networks}.
\newblock In \emph{Proceedings of the 2017 ACM on International Conference on Multimedia Retrieval}, ICMR '17, page 269–277, New York, NY, USA. Association for Computing Machinery.

\bibitem[{Van~der Maaten and Hinton(2008)}]{t-SNE}
Laurens Van~der Maaten and Geoffrey Hinton. 2008.
\newblock Visualizing data using t-sne.
\newblock \emph{Journal of machine learning research}, 9(11).

\bibitem[{Wallace et~al.(2020)Wallace, Stern, and Song}]{Extraction_attack_related_work_2}
Eric Wallace, Mitchell Stern, and Dawn~Xiaodong Song. 2020.
\newblock \href {https://api.semanticscholar.org/CorpusID:216868525} {Imitation attacks and defenses for black-box machine translation systems}.
\newblock In \emph{Conference on Empirical Methods in Natural Language Processing}.

\bibitem[{Wolf et~al.(2020)Wolf, Debut, Sanh, Chaumond, Delangue, Moi, Cistac, Rault, Louf, Funtowicz, Davison, Shleifer, von Platen, Ma, Jernite, Plu, Xu, Le~Scao, Gugger, Drame, Lhoest, and Rush}]{wolf-etal-2020-transformers}
Thomas Wolf, Lysandre Debut, Victor Sanh, Julien Chaumond, Clement Delangue, Anthony Moi, Pierric Cistac, Tim Rault, Remi Louf, Morgan Funtowicz, Joe Davison, Sam Shleifer, Patrick von Platen, Clara Ma, Yacine Jernite, Julien Plu, Canwen Xu, Teven Le~Scao, Sylvain Gugger, Mariama Drame, Quentin Lhoest, and Alexander Rush. 2020.
\newblock \href {https://doi.org/10.18653/v1/2020.emnlp-demos.6} {Transformers: State-of-the-art natural language processing}.
\newblock In \emph{Proceedings of the 2020 Conference on Empirical Methods in Natural Language Processing: System Demonstrations}, pages 38--45, Online. Association for Computational Linguistics.

\bibitem[{Wu et~al.(2020)Wu, Qiao, Chen, Wu, Qi, Lian, Liu, Xie, Gao, Wu, and Zhou}]{MIND-dataset}
Fangzhao Wu, Ying Qiao, Jiun-Hung Chen, Chuhan Wu, Tao Qi, Jianxun Lian, Danyang Liu, Xing Xie, Jianfeng Gao, Winnie Wu, and Ming Zhou. 2020.
\newblock \href {https://doi.org/10.18653/v1/2020.acl-main.331} {{MIND}: A large-scale dataset for news recommendation}.
\newblock In \emph{Proceedings of the 58th Annual Meeting of the Association for Computational Linguistics}, pages 3597--3606, Online. Association for Computational Linguistics.

\bibitem[{Xu and He(2023)}]{EMNLP-2023-security-tutorial}
Qiongkai Xu and Xuanli He. 2023.
\newblock \href {https://doi.org/10.18653/v1/2023.emnlp-tutorial.2} {Security challenges in natural language processing models}.
\newblock In \emph{Proceedings of the 2023 Conference on Empirical Methods in Natural Language Processing: Tutorial Abstracts}, pages 7--12, Singapore. Association for Computational Linguistics.

\bibitem[{Xu et~al.(2022)Xu, He, Lyu, Qu, and Haffari}]{xu-etal-2022-student}
Qiongkai Xu, Xuanli He, Lingjuan Lyu, Lizhen Qu, and Gholamreza Haffari. 2022.
\newblock \href {https://aclanthology.org/2022.coling-1.251} {Student surpasses teacher: Imitation attack for black-box {NLP} {API}s}.
\newblock In \emph{Proceedings of the 29th International Conference on Computational Linguistics}, pages 2849--2860, Gyeongju, Republic of Korea. International Committee on Computational Linguistics.

\bibitem[{Yue et~al.(2021)Yue, He, Zeng, and McAuley}]{Extraction_attack_related_work}
Zhenrui Yue, Zhankui He, Huimin Zeng, and Julian McAuley. 2021.
\newblock \href {https://doi.org/10.1145/3460231.3474275} {Black-box attacks on sequential recommenders via data-free model extraction}.
\newblock In \emph{Proceedings of the 15th ACM Conference on Recommender Systems}, RecSys '21, page 44–54, New York, NY, USA. Association for Computing Machinery.

\bibitem[{Zhang et~al.(2015)Zhang, Zhao, and LeCun}]{ag_news-dataset}
Xiang Zhang, Junbo Zhao, and Yann LeCun. 2015.
\newblock Character-level convolutional networks for text classification.
\newblock In \emph{Proceedings of the 28th International Conference on Neural Information Processing Systems - Volume 1}, NIPS'15, page 649–657, Cambridge, MA, USA. MIT Press.

\bibitem[{Zhang et~al.(2023)Zhang, Xiao, Li, Lv, Qi, Liu, Wang, Jiang, and Sun}]{Backdoor_neuroBA}
Zhengyan Zhang, Guangxuan Xiao, Yongwei Li, Tian Lv, Fanchao Qi, Zhiyuan Liu, Yasheng Wang, Xin Jiang, and Maosong Sun. 2023.
\newblock \href {https://doi.org/10.1007/s11633-022-1377-5} {Red alarm for pre-trained models: Universal vulnerability to neuron-level backdoor attacks}.
\newblock \emph{Machine Intelligence Research}, 20(2):180--193.

\end{thebibliography}

\clearpage
\appendix
\section*{Appendix}
\section{Experimental Settings}
\label{appendix:exp-setting}
We leverage the standard codebase of the Transformers \citep{wolf-etal-2020-transformers} library and AdamW \citep{loshchilov2018decoupled} algorithm for model training and development. Likewise, we use scikit-learn \citep{scikit-learn} for clustering algorithms and other utility calculations. We use GPT-3 text-embedding-002 API as original benign embeddings and the BERT \citep{devlin-etal-2019-bert} model as the victim model.
We perform all the experiments on a single A100 GPU with CUDA 11.7 and pytorch 2.1.2. We assume that both the victim model and imitators use the same datasets to separate the effects of the watermarking technique from other factors. Furthermore, we assume that the extracted model is trained only using the watermarked outputs from the victim model.
Finally, we implement the \prevWM and other experiments following their default configurations and settings, \ie, $m=4, n=20, \text{and frequency interval} = [0.5\%, 1\%]$. The only exception is the $R=10$ case in \ourdefence, where we use $n=50$ to have enough trigger words. A standard dataset, WikiText~\cite{WikiText-dataset} consisting of $1,801,350$ entries, serves as a hold-out dataset for selecting moderate-frequency words as watermark triggers ($T$). 

\section{Similarity Distribution Plots}
\label{appendix:other-cos-sim-dist-plots}
\begin{figure*}[h]
     \centering
     \begin{subfigure}[b]{0.25\textwidth}
         \centering
         \includegraphics[width=\textwidth]{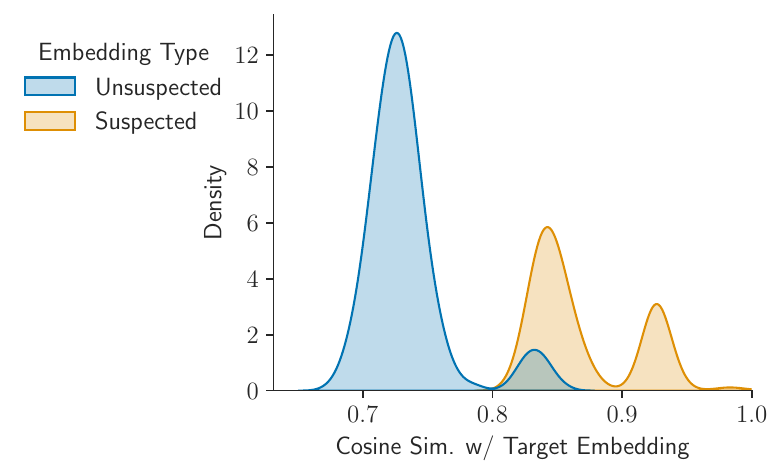}
         \caption{\sst}
         \label{sst2}
     \end{subfigure}
     \begin{subfigure}[b]{0.23\textwidth}
         \centering
         \includegraphics[width=\textwidth]{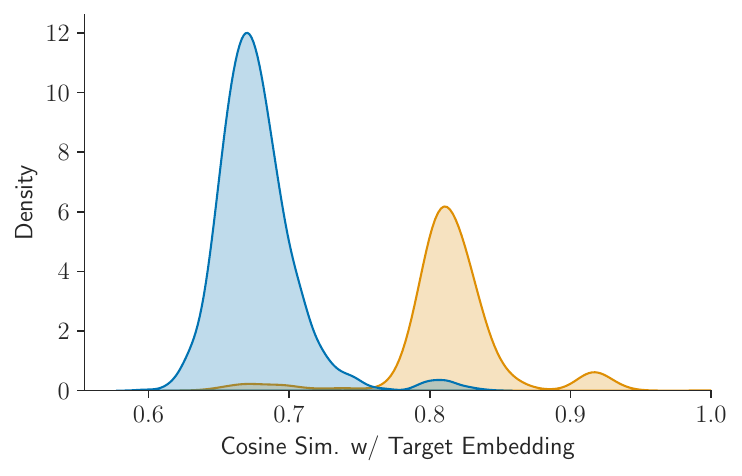}
         \caption{\mind}
         \label{mind}
     \end{subfigure}
     \begin{subfigure}[b]{0.23\textwidth}
         \centering
         \includegraphics[width=\textwidth]{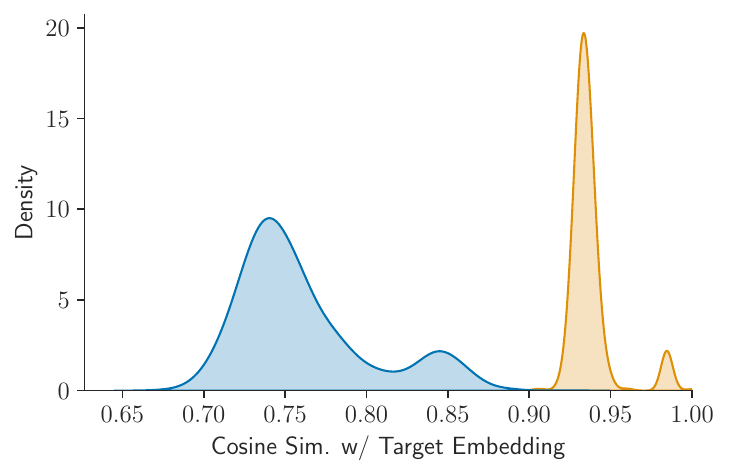}
         \caption{\agnews}
         \label{agnews}
     \end{subfigure}
     \begin{subfigure}[b]{0.23\textwidth}
         \centering
         \includegraphics[width=\textwidth]{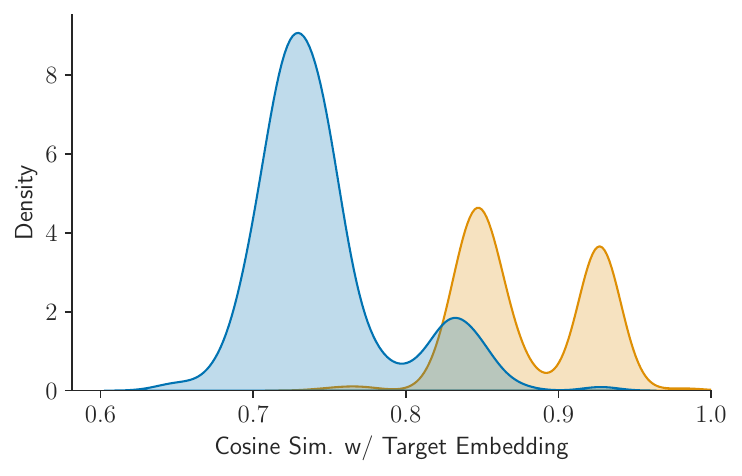}
         \caption{\enron}
         \label{enron}
     \end{subfigure}
        \caption{Distribution plots for cosine similarities between different types of embeddings and the target embedding for different datasets.
        }
        \label{fig:other-cos-sim-dist-sim-with-target-emb}
\end{figure*}
The observations (captured in ~\reffig{fig:other-cos-sim-dist-sim-with-target-emb}) for other datasets are similar to \sst as seen in \reffig{fig:sst2-cos-sim-dist-sim-with-target-emb}, \ie, watermarked embeddings are closer to target embedding, and there is a clear difference in similarities for watermarked and non-watermarked embeddings. Due to skewness between the number of suspected and unsuspected embeddings, we employ sampling for unsuspected entries in these plots.

\section{\ourattack Attack Analyses}
\label{appendix:cse-ablation}
In this section, we perform detailed ablation studies for \ourattack attack.
\subsection{Comparison of Clustering Algorithms}
\label{appendix:clustering-algo-comparision}
\begin{table}[h]
    \begin{minipage}{\columnwidth}
    \resizebox{\textwidth}{!}{%
    \begin{tabular}{cccccc}
        \toprule
        \multirow{2}{*}{Dataset} & \multicolumn{3}{c}{Detection Performance} \\
        \cmidrule(lr){2-4}
        {} & {p-value} & {$\Delta_{cos}(\%)$} & {$\Delta_{l2}(\%)$}\\
        \toprule
        \sst & $> 0.02$ & 1.00$\pm$0.40 & -2.00$\pm$0.80 \\
        \mind & $> 0.55$ & 0.28$\pm$0.31 & -0.55$\pm$0.63 \\
        \agnews & $> 0.10$ & 0.45$\pm$0.42 & -0.90$\pm$0.84 \\
        \enron & $> 0.56$ & 0.23$\pm$0.52 & -0.47$\pm$1.04 \\
        \bottomrule
    \end{tabular}}
    \caption{\ourattack performance using GMM clustering algorithm, similar to \kmeans algorithm (tabulated in \reftab{table:attack-performance}).}
    \label{comparison of Clustering algo}
    \end{minipage}
\end{table}

The previous experiments utilize \kmeans as the clustering algorithm. However, alternative algorithms such as Gaussian Mixture Models (\gmm) \cite{GMM} are also valid options. The subsequent table, \reftab{comparison of Clustering algo}, illustrates the comparative performance between \kmeans and \gmm. While \kmeans exhibits superior performance in downstream utility for \agnews, \enron, and \mind datasets, it is less confident for delta values in the case of \enron and \mind. Overall both algorithms demonstrate satisfactory performance, suggesting that the clustering module for \ourattack is universally adaptable to different clustering algorithms.

\subsection{Number of Clusters ($n$)}
\label{appendix:num-clusters-cse}
\begin{figure*}[h]
    \centering
    \begin{subfigure}{0.23\textwidth}
        \centering
        \includegraphics[width=\linewidth]{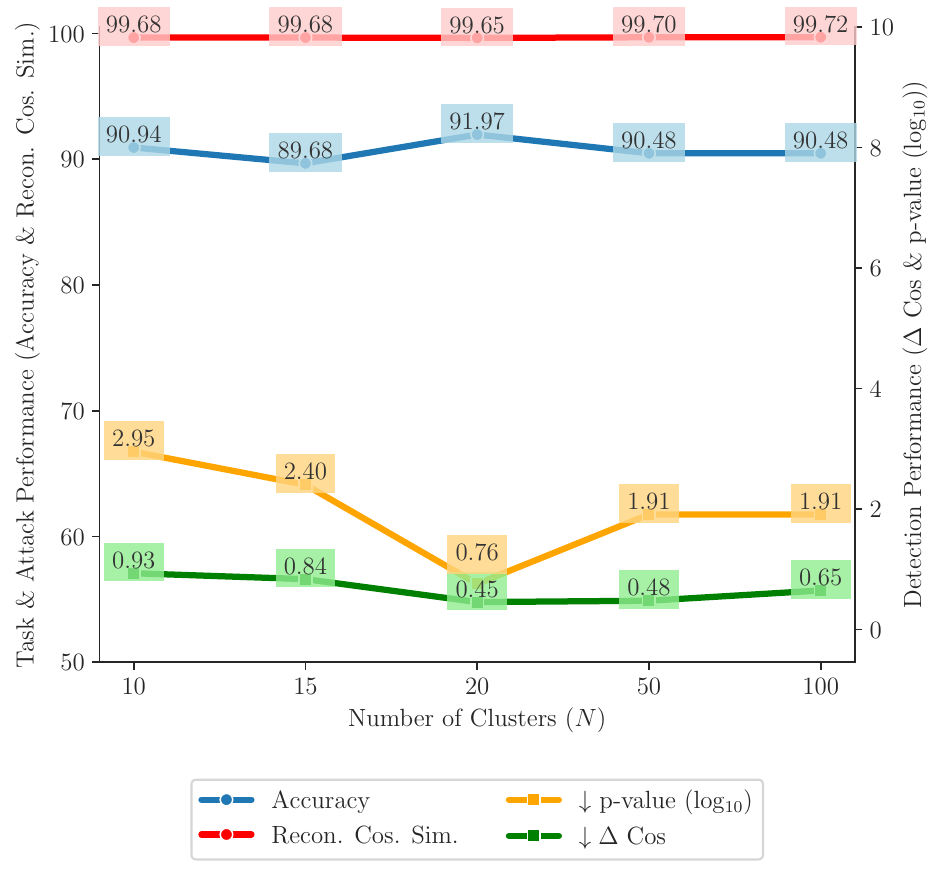}
        \caption{\sst}
    \end{subfigure}
    \begin{subfigure}{0.23\textwidth}
        \centering
        \includegraphics[width=\linewidth]{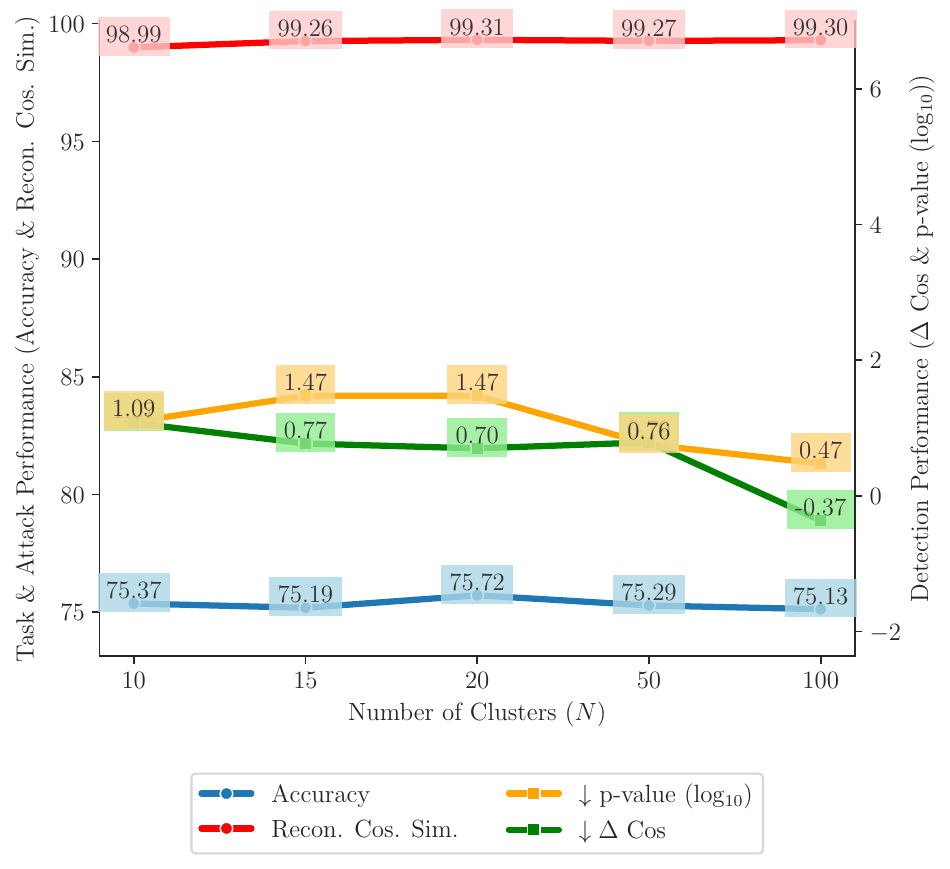}
        \caption{\mind}
    \end{subfigure}
    \begin{subfigure}{0.23\textwidth}
        \centering
        \includegraphics[width=\linewidth,keepaspectratio]{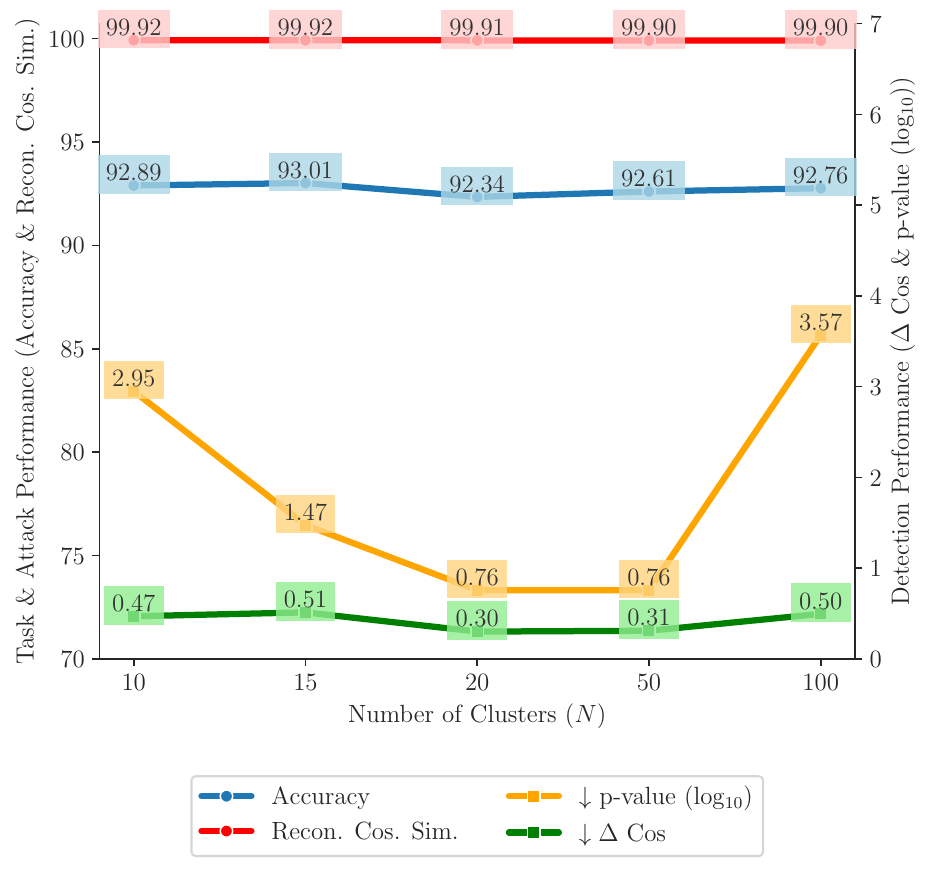}
        \caption{\agnews}
    \end{subfigure}
    \begin{subfigure}{0.23\textwidth}
        \centering
        \includegraphics[width=\linewidth,keepaspectratio]{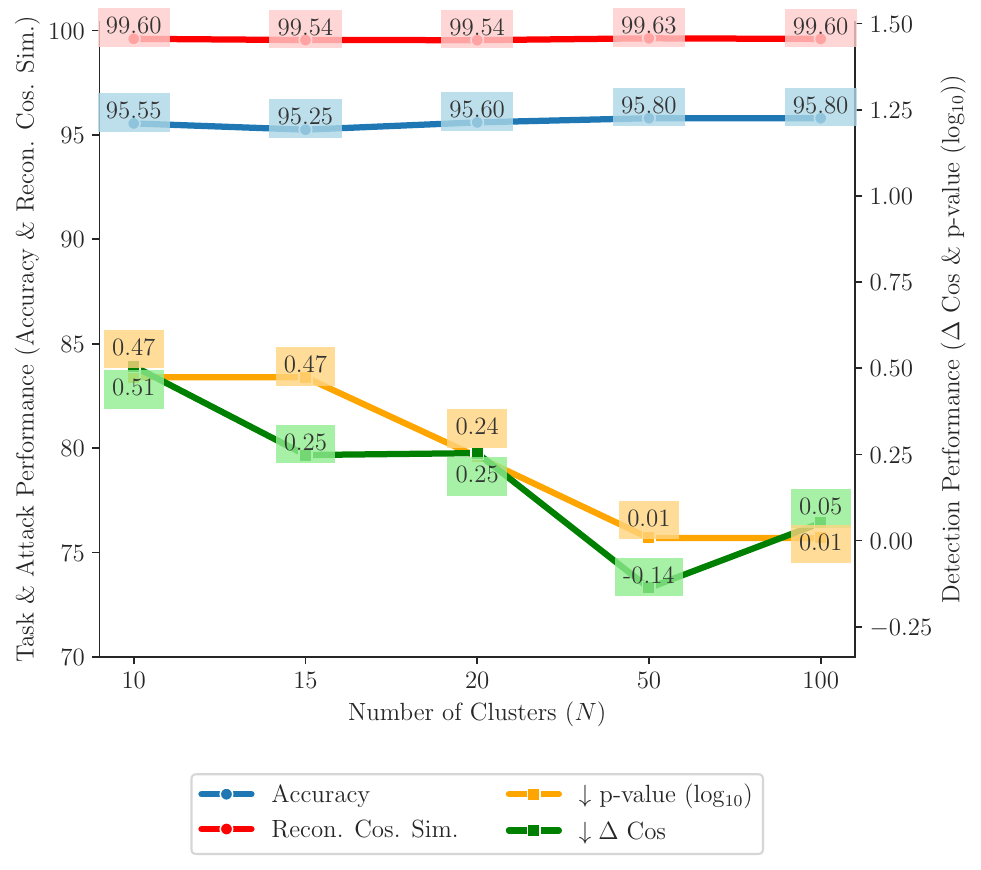}
        \caption{\enron}
    \end{subfigure}
    \caption{The impact of cluster numbers ($n$) in \ourattack for different datasets.}
    \label{fig:CSE-cluster-num}
\end{figure*}

\begin{figure*}[h]
     \centering
     \begin{subfigure}[b]{0.10\textwidth}
         \centering
         \includegraphics[width=\textwidth]{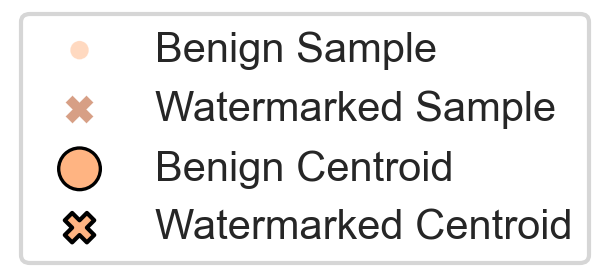}
     \end{subfigure}
     \hfill
     \begin{subfigure}[b]{0.20\textwidth}
         \centering
         \includegraphics[width=\textwidth]{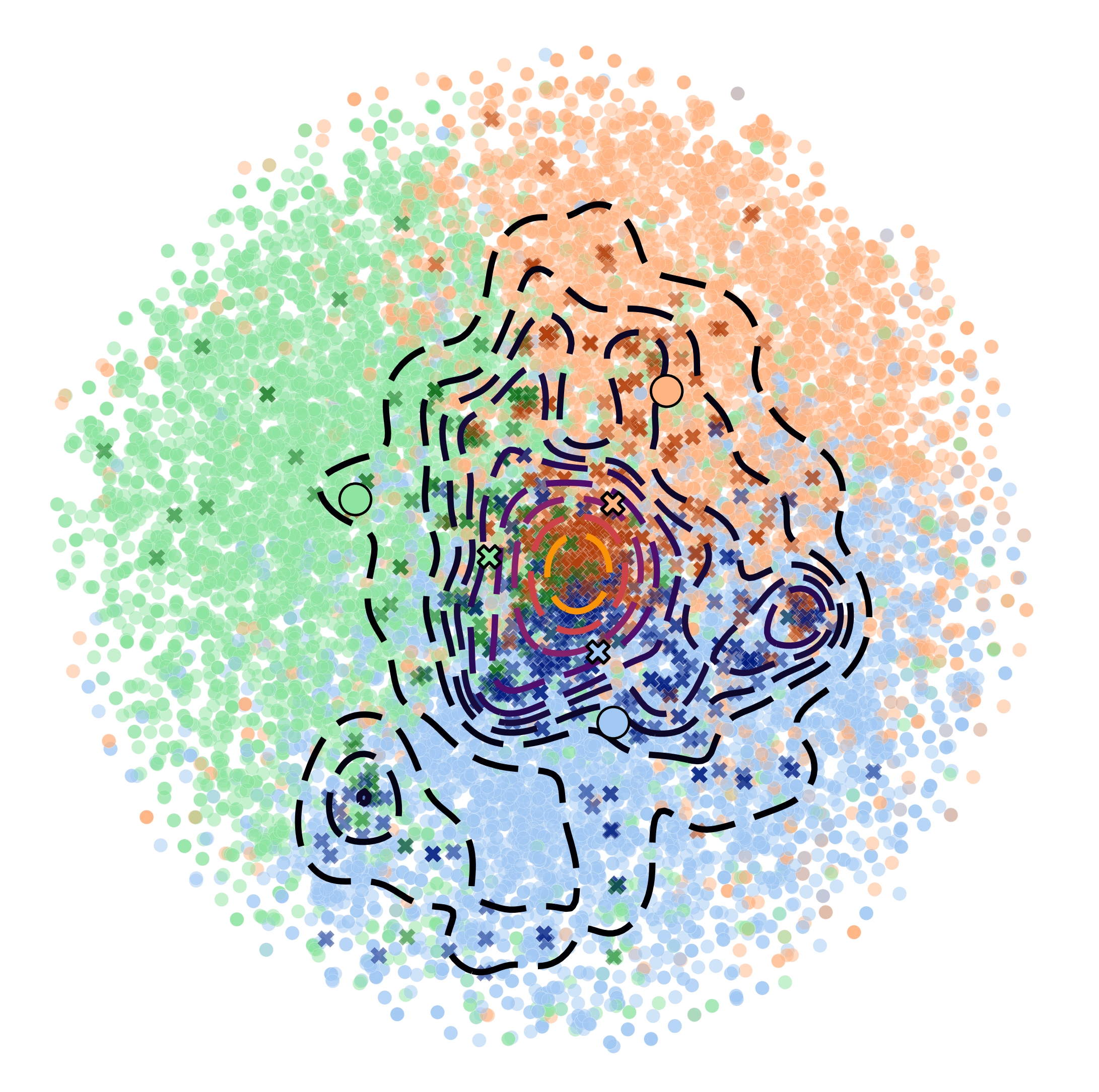}
         \caption{\sst}
         \label{sst2-clustering}
     \end{subfigure}
     \begin{subfigure}[b]{0.20\textwidth}
         \centering
         \includegraphics[width=\textwidth]{asset/imgs/clustering-contour/clustering-contour-improved/cluster-mind-kmeans-3-clusters-seed-47-clustering-plot-t-sne-both-centroids-_6,_6_.png}
         \caption{\mind}
         \label{mind-clustering}
     \end{subfigure}
     \hfill
     \begin{subfigure}[b]{0.20\textwidth}
         \centering
         \includegraphics[width=\textwidth]{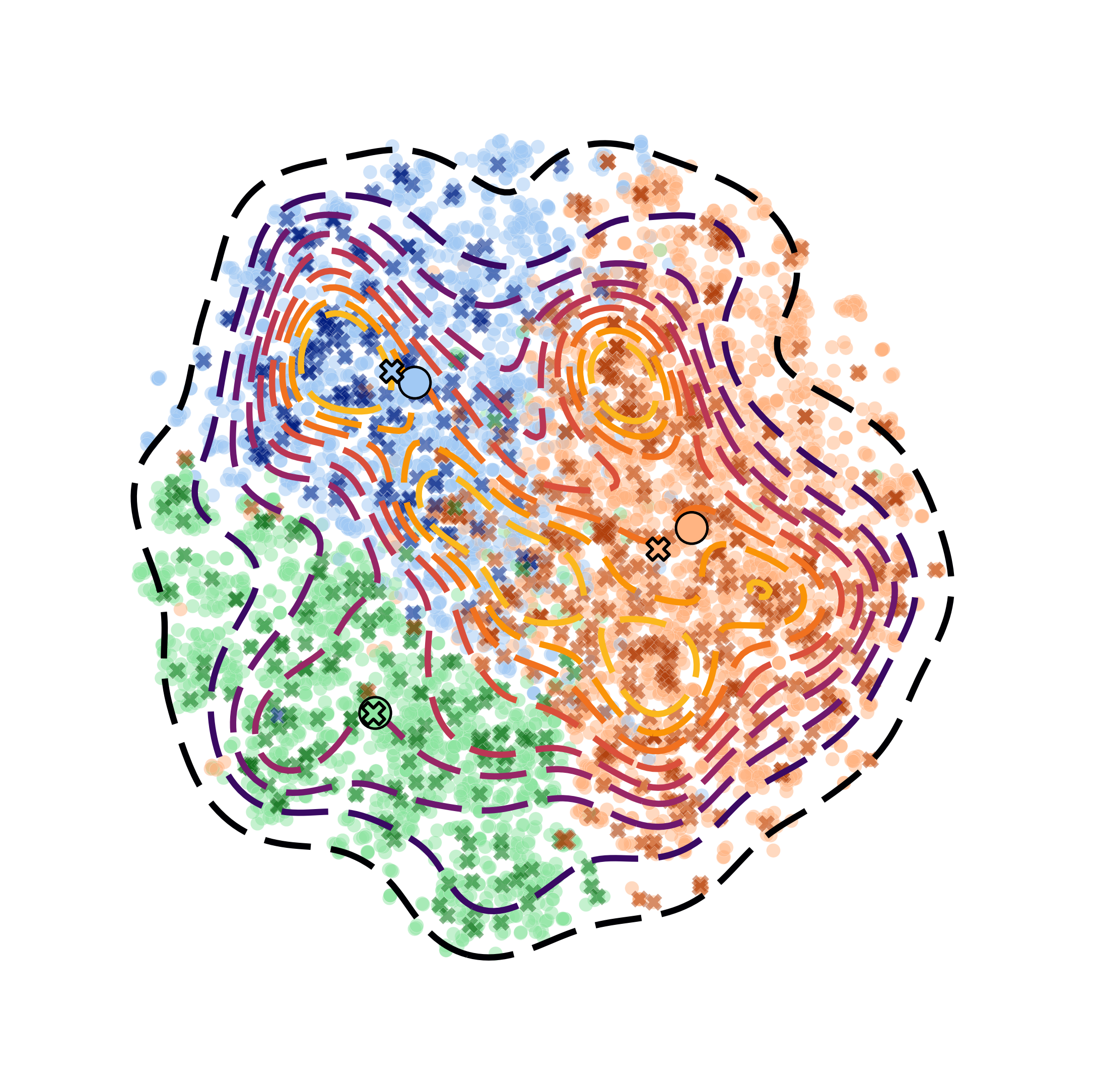}
         \caption{\agnews}
         \label{ag_news-clustering}
     \end{subfigure}
     \hfill
     \begin{subfigure}[b]{0.20\textwidth}
         \centering
         \includegraphics[width=\textwidth]{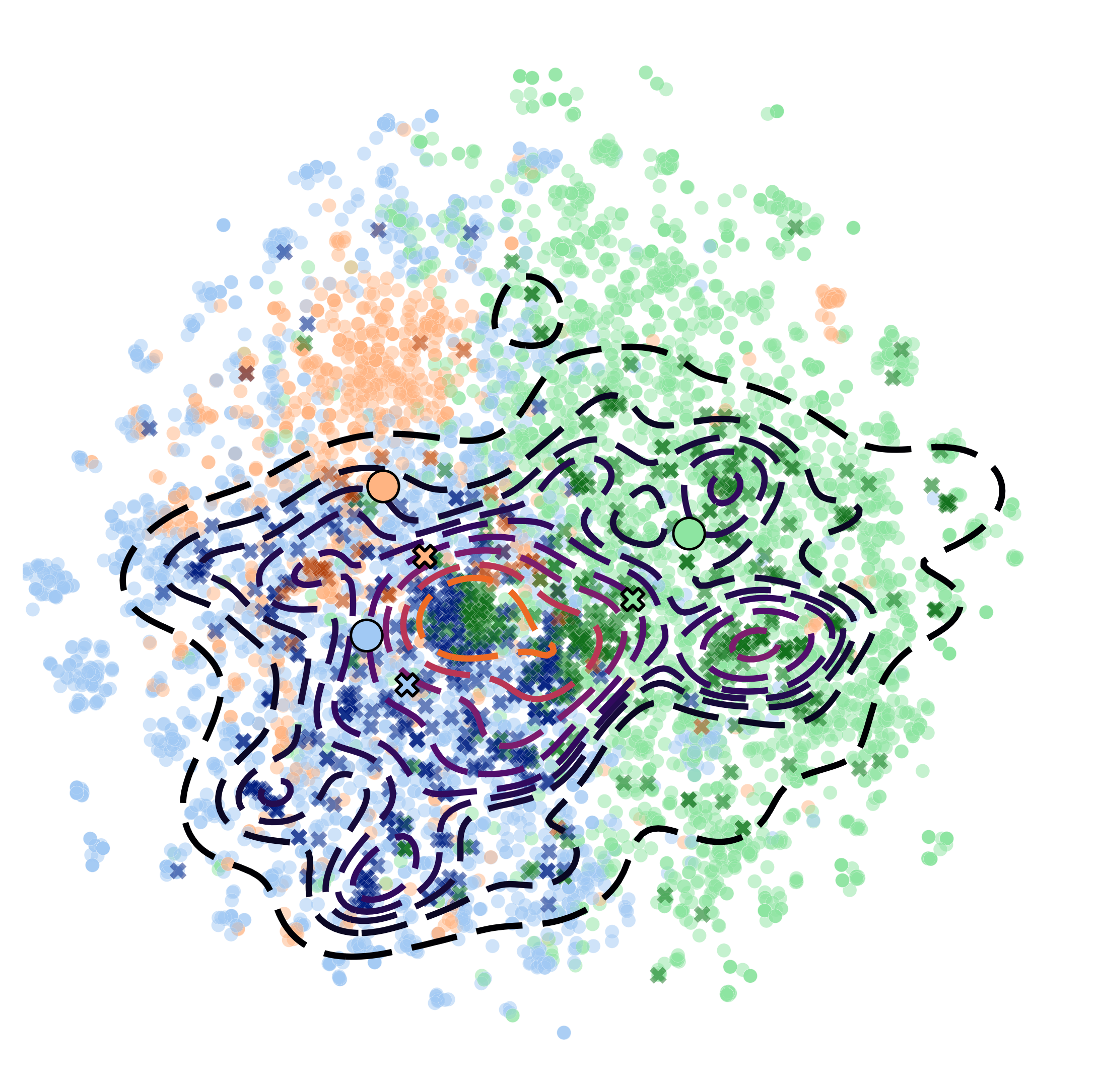}
         \caption{\enron}
         \label{enron-clustering}
     \end{subfigure}
        \caption{t-SNE \citep{t-SNE} visualisations for \kmeans clustering ($n=3$) for different datasets. It is evident from the plots that the watermarked samples are not clustered together but instead spread across the embedding space with non-coinciding centroids.}
        \label{fig:clustering-with-contour-plots}
\end{figure*}

We can see from \reffig{fig:CSE-cluster-num} that there is no significant role in the number of clusters ($n$). In all the cases, \ourattack attack is successful, though we use $n=20$ in our experiments. However, considering pairwise distance comparison, it is preferable to have fewer clusters to increase the likelihood of watermarked pairs. We also visualize these clusters in \reffig{fig:clustering-with-contour-plots}. 

\subsection{Number of Principal Components ($K$)}

\begin{figure*}[h]
    \centering
    \begin{subfigure}{0.22\textwidth}
        \centering    \includegraphics[width=\linewidth,keepaspectratio]{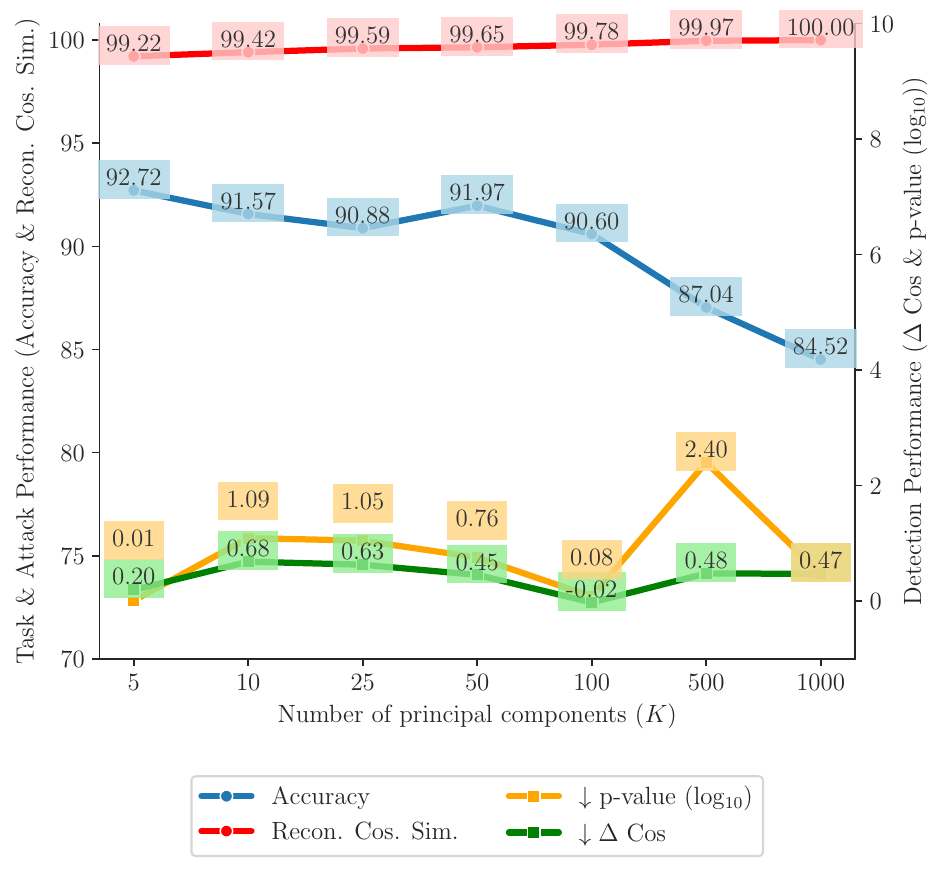}
    \caption{\sst}
    \end{subfigure}
    \begin{subfigure}{0.22\textwidth}
        \centering
        \includegraphics[width=\linewidth]{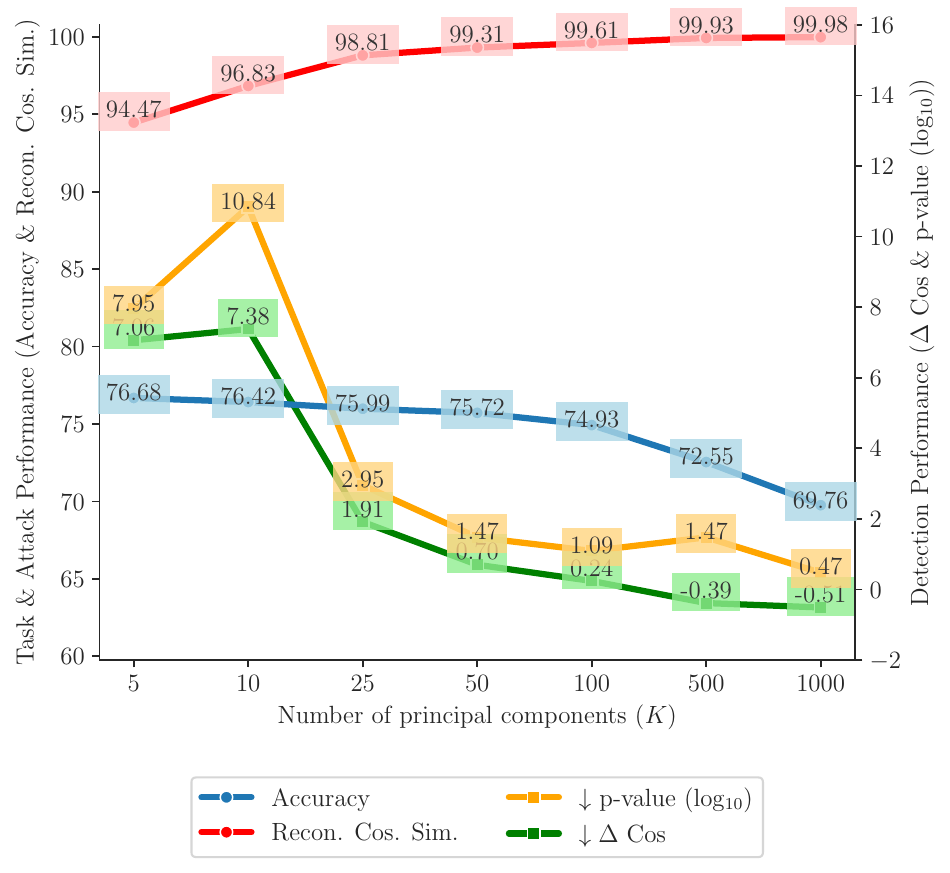}
        \caption{\mind}
        
    \end{subfigure}
    \begin{subfigure}{0.22\textwidth}
        \centering
        \includegraphics[width=\linewidth,keepaspectratio]{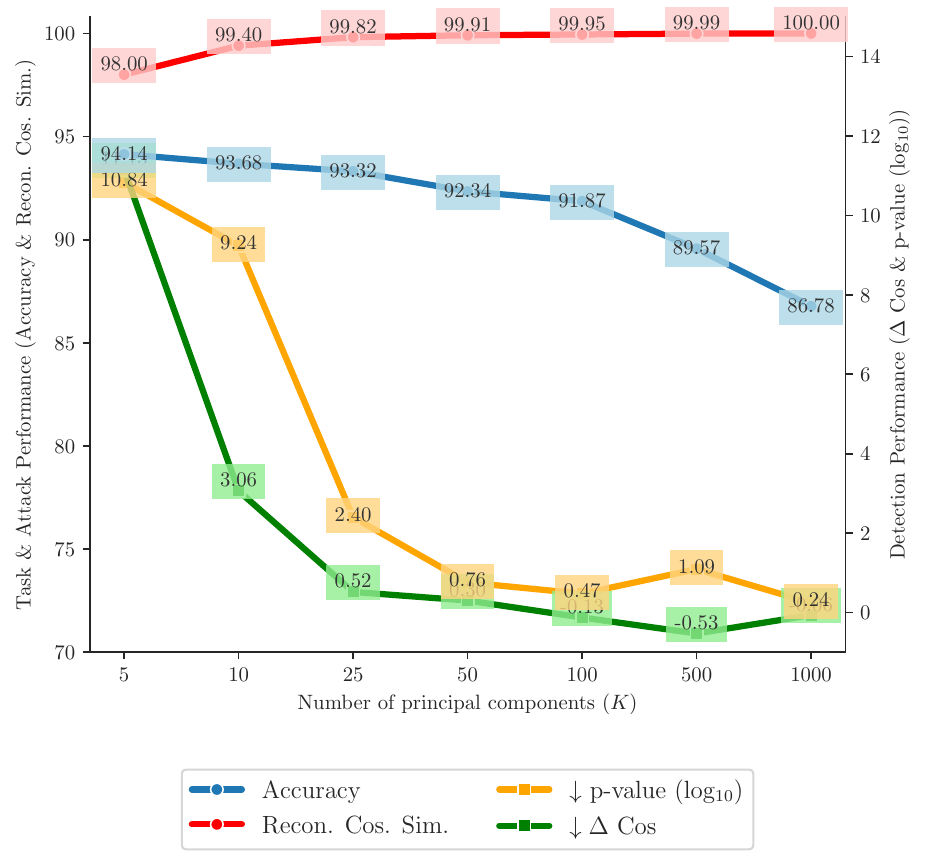}
        \caption{\agnews}
    \end{subfigure}
    \begin{subfigure}{0.22\textwidth}
        \centering
        \includegraphics[width=\linewidth,keepaspectratio]{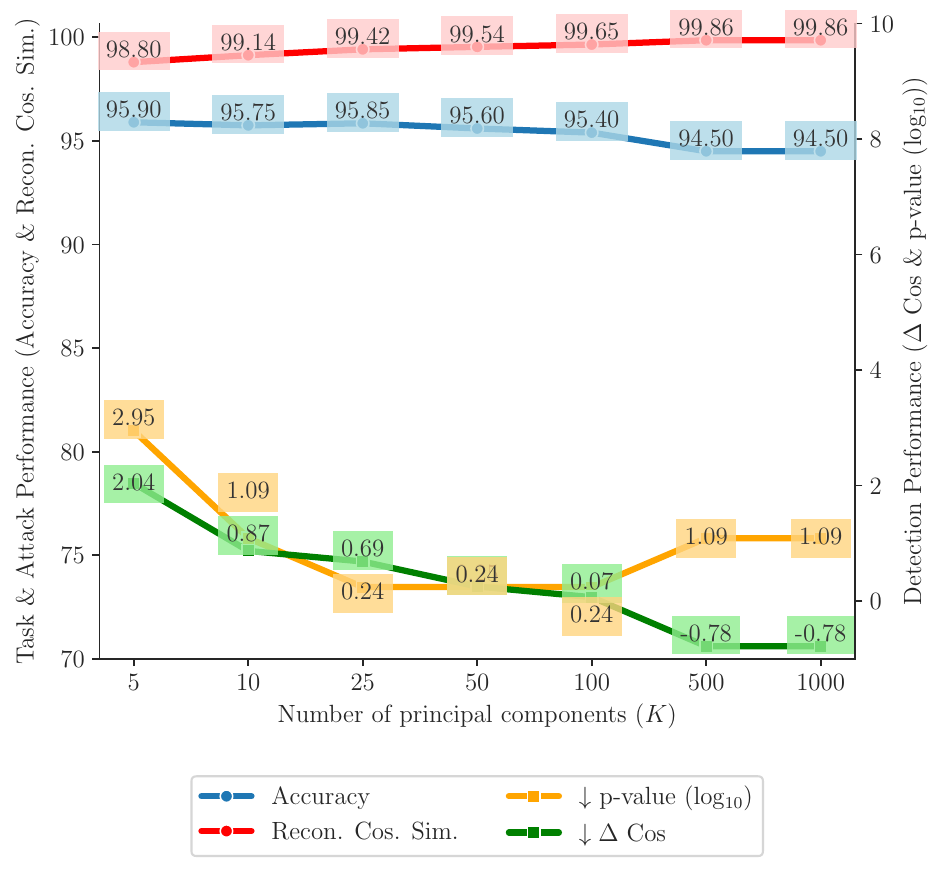}
        \caption{\enron}
    \end{subfigure}
    \caption{The impact of number of principal components ($N$) in \ourattack for different datasets.}
    \label{fig:CSE-pc}
\end{figure*}

In \refsec{sec:target-emb-reconstruction}, we formulated an optimization problem to compute how much of the watermark we are recovering for a given number of principal components ($\vc_k$). 
As expected, with increasing $K$, we will recover more of the target embeddings, as noted from the red line in \reffig{fig:CSE-pc}. However, the utility metrics deteriorate more significantly (blue line). Meanwhile, a lower $K$ does not recover enough target embedding to bypass copyright verification (yellow and green lines -- detection performance). To achieve the best of both worlds—downstream utility and avoiding watermark—we must strike a balance, and 50 components seem appropriate.
Further, in some cases, we observe that the increasing $K$ does not affect the downstream utility metrics. The downstream task's simplicity could be the cause of this. For example, \enron dataset is a binary classification task wherein required data could be represented in a few embedding dimensions.

\subsection{Impact of Attacker Model Size}

\begin{table}[h]
\begin{minipage}{0.95\columnwidth}
\resizebox{\textwidth}{!}{%
    \centering
    \begin{tabular}{ccccc}
    \toprule
    \multirow{2}{*}{Dataset} & \multirow{2}{*}{Size} & \multicolumn{3}{c}{Detection Performance} \\
     \cmidrule(lr){3-5}
    {}  & {} & p-value & {$\Delta_{cos}(\%)$} & {$\Delta_{l2}(\%)$}\\
    \toprule
        \sst &  \multirow{4}{*}{Small} & $>0.57$ & 0.41 & -0.81 \\
        \mind &  & $> 10^{-4}$ & 1.38 & -2.76 \\
        \agnews &  & $>0.5$7 & -0.08 & 0.17 \\
        \enron &   & $>0.08$ & 0.63 & -1.27 \\
        \midrule
        \sst & \multirow{4}{*}{Base}  & $>0.17$ & 0.00 &-0.01 \\ 
        \mind &  & $>10^{-3}$ & -0.01 & 0.03 \\
        \agnews &   & $>0.17$ & 0.00 &-0.01 \\
        \enron &   &$>0.57$ & 0.00 & -0.01 \\ 
        \midrule
        \sst & \multirow{4}{*}{Large}  & $>0.56$ & 0.89 & -1.79 \\
        \mind &  & $>0.17$ & 0.37 & -0.74 \\
        \agnews & & $>0.98$ & 0.04 & -0.09 \\
        \enron &  & $>0.57$ & 0.28 & -0.56 \\
    \bottomrule
    \end{tabular}}
\caption{The impact of extracted model size on \ourattack performance.}
\label{table:attack-model-sizes}
\end{minipage}
\end{table}

We evaluate if there are any differences in our attack's performance for different attacker model sizes. This is tested by performing experiments on small, base, and large versions of the BERT \citep{devlin-etal-2019-bert} model. As illustrated in the \reftab{table:attack-model-sizes}, the attack effectively bypasses the watermark when the stealer uses different sizes of the backbone model.

\section{\ourdefence Defense Analyses}
\label{appendix:ablation-our-defence}

\subsection{Remaining Dataset Results}
\label{warden-other-dataset-results}
\begin{figure*}[h]
    \centering
    \begin{subfigure}{0.23\textwidth}
        \centering
        \includegraphics[width=\linewidth]{asset/imgs/WARDEN/sst2_v2.pdf}
        \caption{\sst}
        \label{fig:sst2}
    \end{subfigure}
    \begin{subfigure}{0.23\textwidth}
        \centering
        \includegraphics[width=\linewidth]{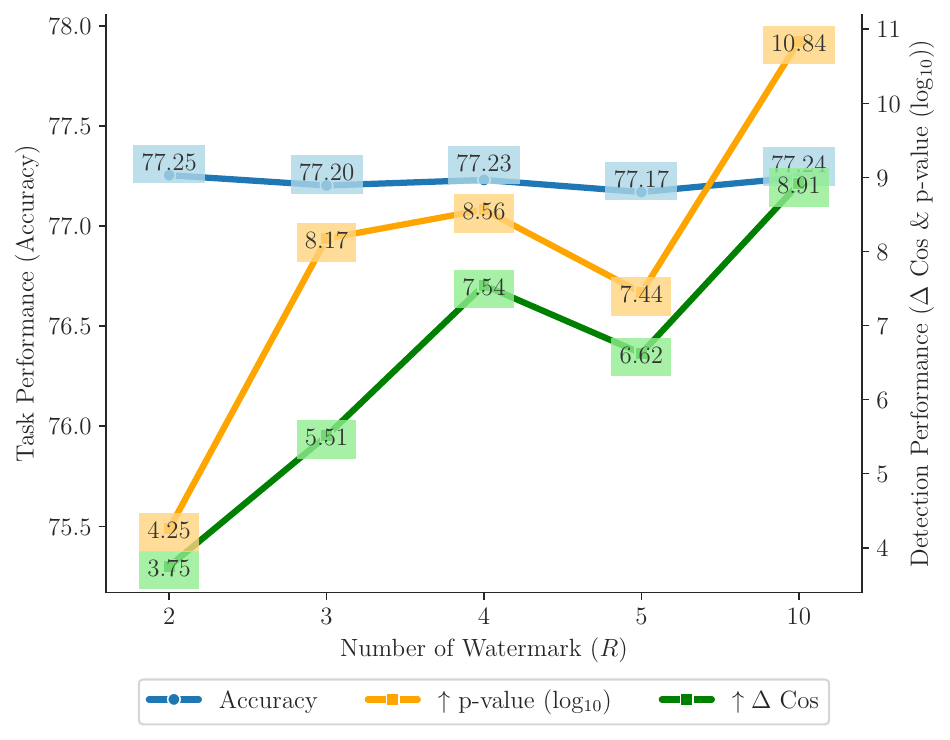}
        \caption{\mind}
        \label{fig:mind}
    \end{subfigure}
    \begin{subfigure}{0.23\textwidth}
        \centering
        \includegraphics[width=\linewidth,keepaspectratio]{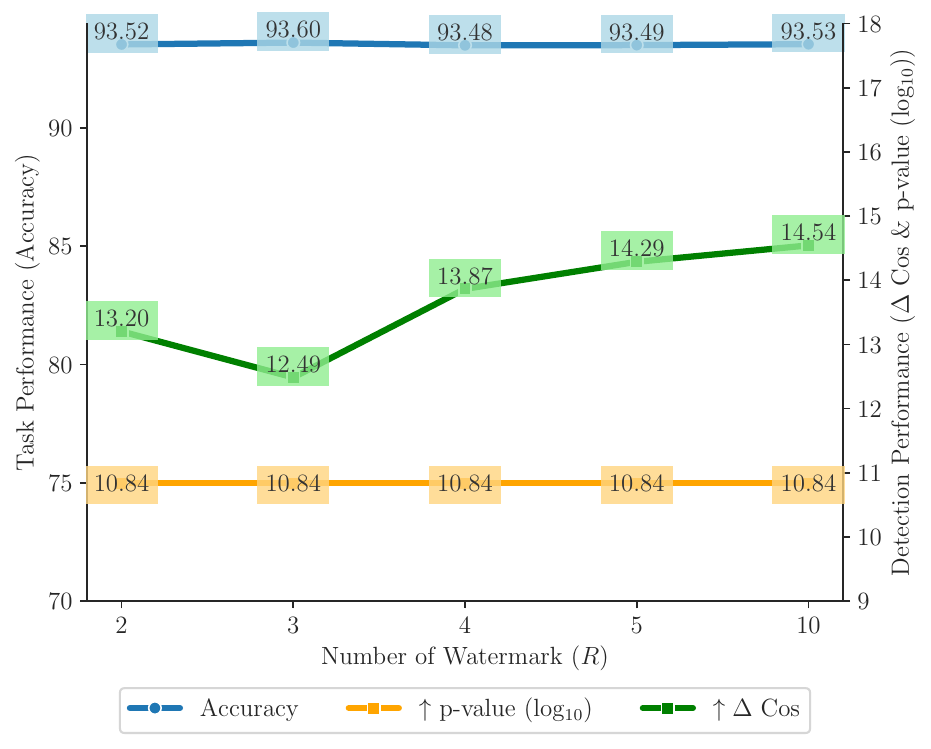}
        \caption{\agnews}
        \label{fig:agnews}
    \end{subfigure}
    \begin{subfigure}{0.23\textwidth}
        \centering
        \includegraphics[width=\linewidth,keepaspectratio]{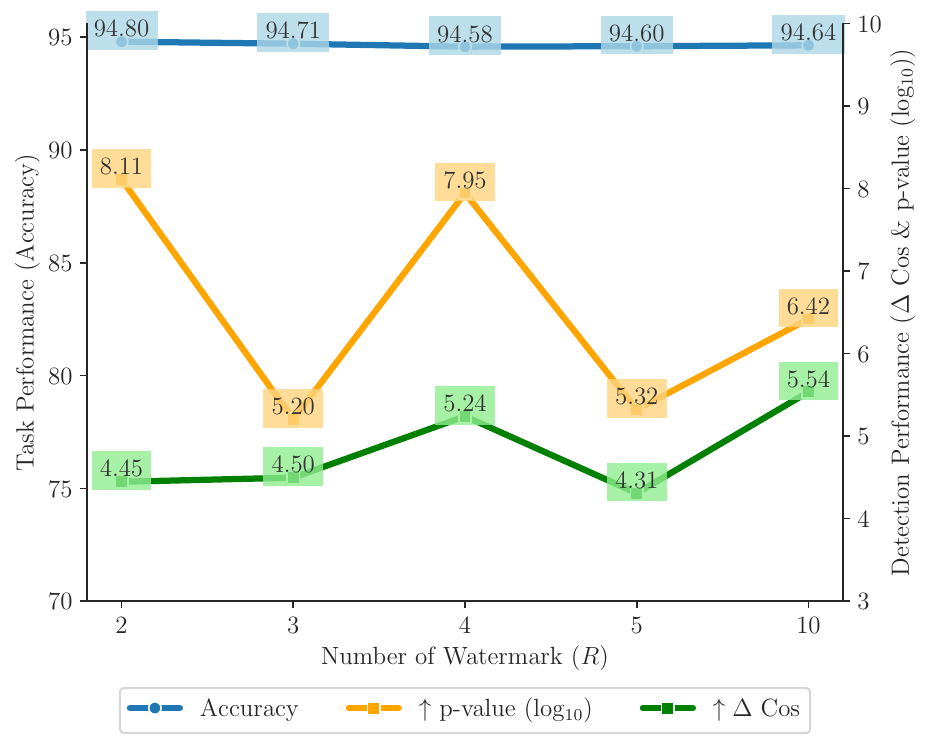}
        \caption{\enron}
        \label{fig:enron}
    \end{subfigure}
    \caption{The impact of number of watermarks ($R$) in \ourdefence for different datasets. As expected, detection performance (yellow and green lines) shows an upward trend with stable task performance.}
    \label{fig:num-watermarks-warden}
\end{figure*}

\begin{figure*}[h]
    \centering
    \begin{subfigure}{0.23\textwidth}
        \centering
        \includegraphics[width=\linewidth]{asset/imgs/CSE-on-WARDEN/sst2_v2.pdf}
        \caption{\sst}
    \end{subfigure}
    \begin{subfigure}{0.23\textwidth}
        \centering
        \includegraphics[width=\linewidth]{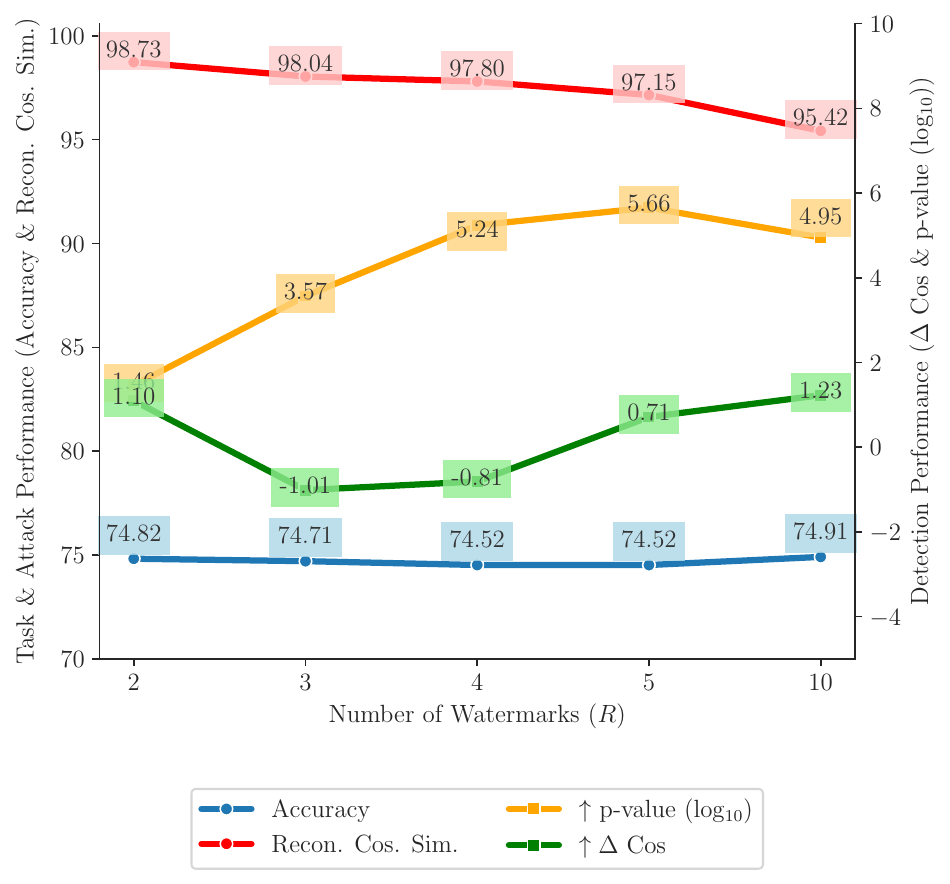}
        \caption{\mind}
    \end{subfigure}
    \begin{subfigure}{0.23\textwidth}
        \centering
        \includegraphics[width=\linewidth,keepaspectratio]{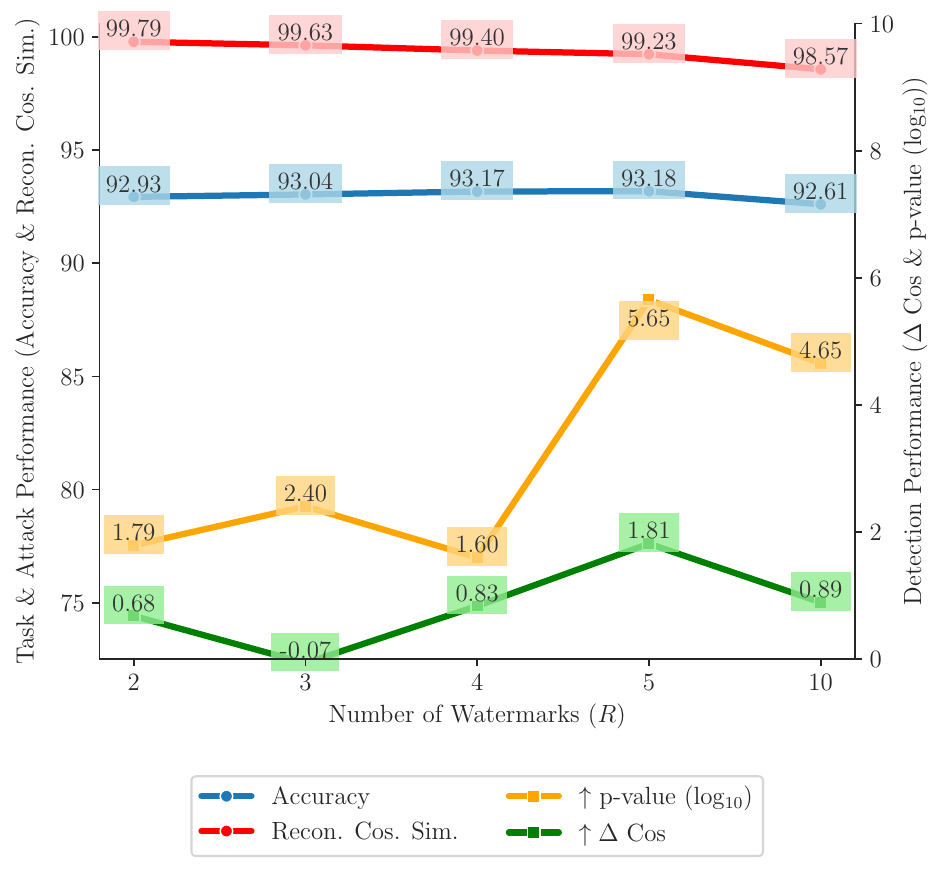}
        \caption{\agnews}
    \end{subfigure}
    \begin{subfigure}{0.23\textwidth}
        \centering
        \includegraphics[width=\linewidth,keepaspectratio]{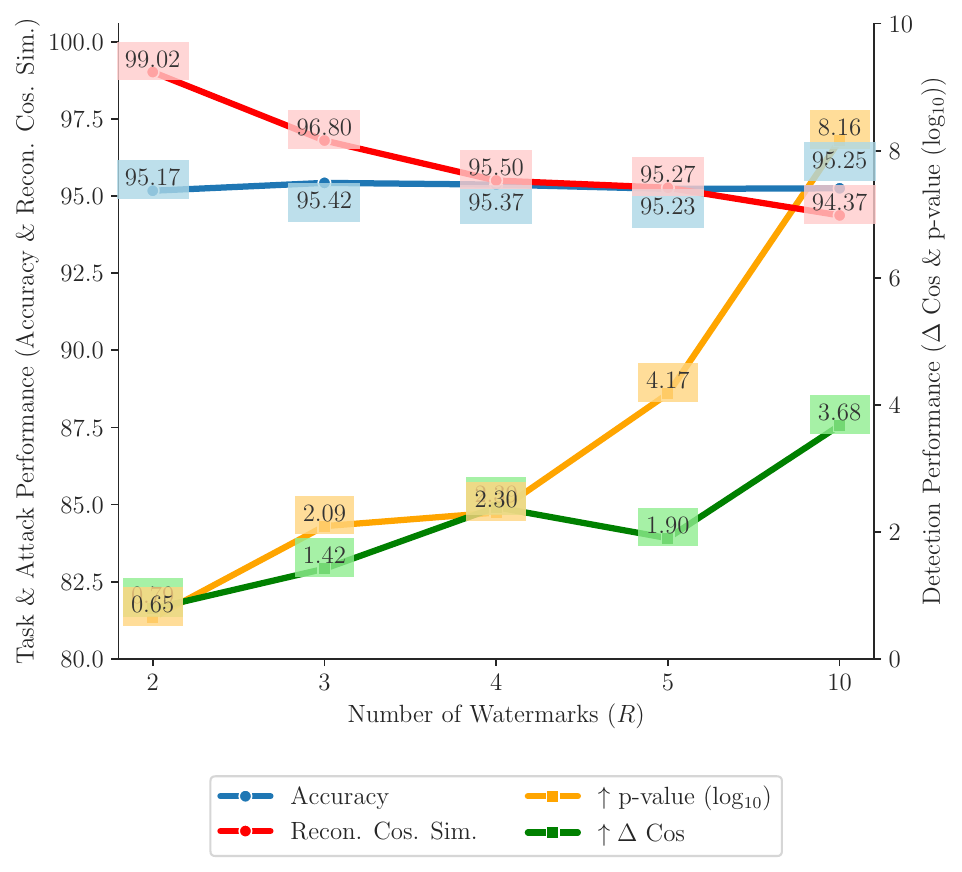}
        \caption{\enron}
    \end{subfigure}
    \caption{The impact of number of watermarks ($R$) in \ourdefence against \ourattack for different datasets. The observation is similar to \reffig{fig:num-watermarks-warden}, along with a decreasing trend in attack performance (red line) demonstrating the effectiveness of \ourdefence defense against \ourattack attack.}
    \label{fig:num-watermarks-warden-cse}
\end{figure*}

\begin{figure*}
    \centering
    \begin{subfigure}{0.22\textwidth}
        \centering
        \includegraphics[width=\linewidth]{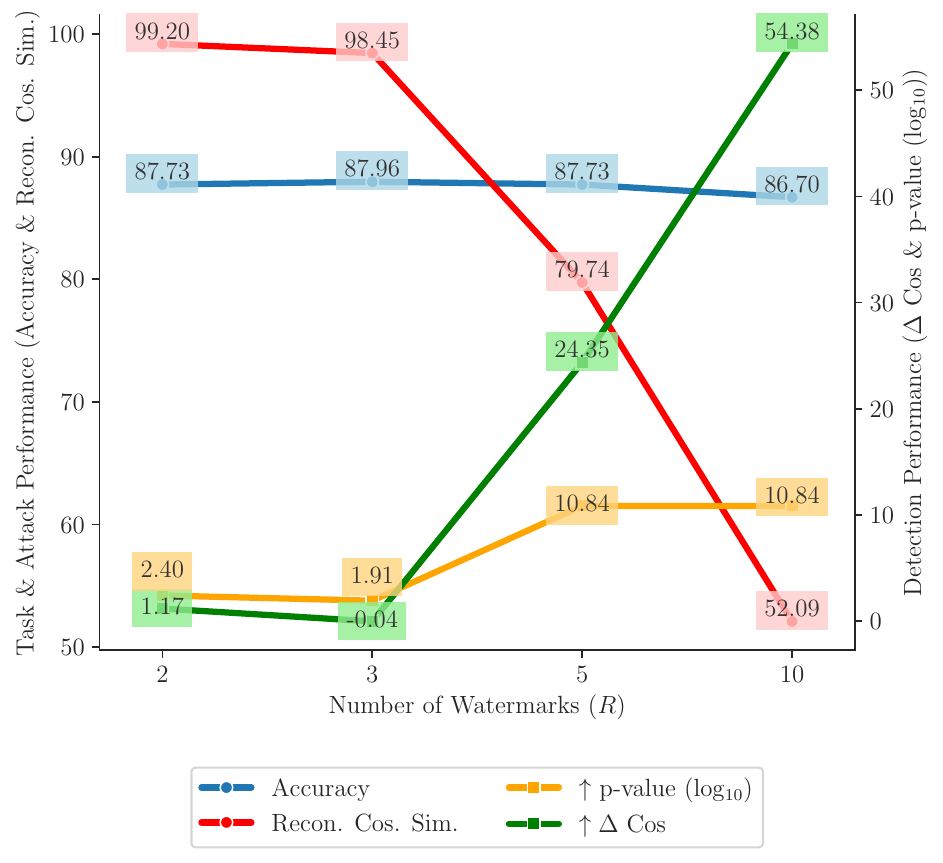}
        \caption{\sst}
    \end{subfigure}
    \begin{subfigure}{0.22\textwidth}
        \centering
        \includegraphics[width=\linewidth]{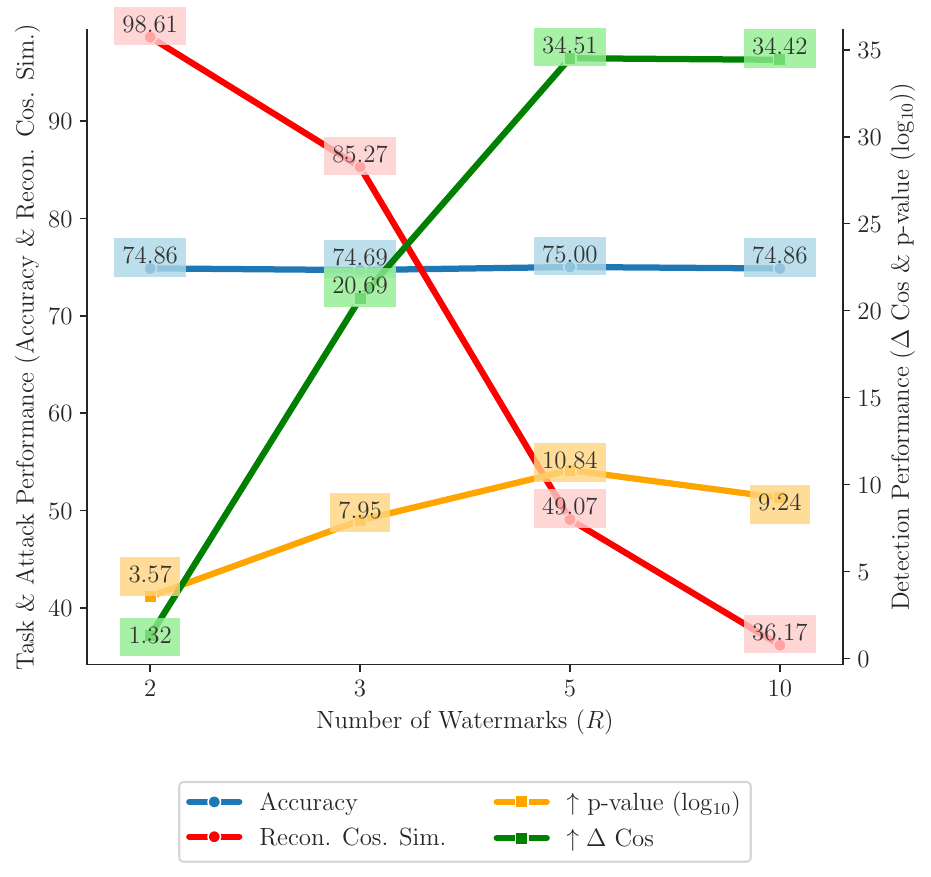}
        \caption{\mind}
    \end{subfigure}
    \begin{subfigure}{0.22\textwidth}
        \centering
        \includegraphics[width=\linewidth,keepaspectratio]{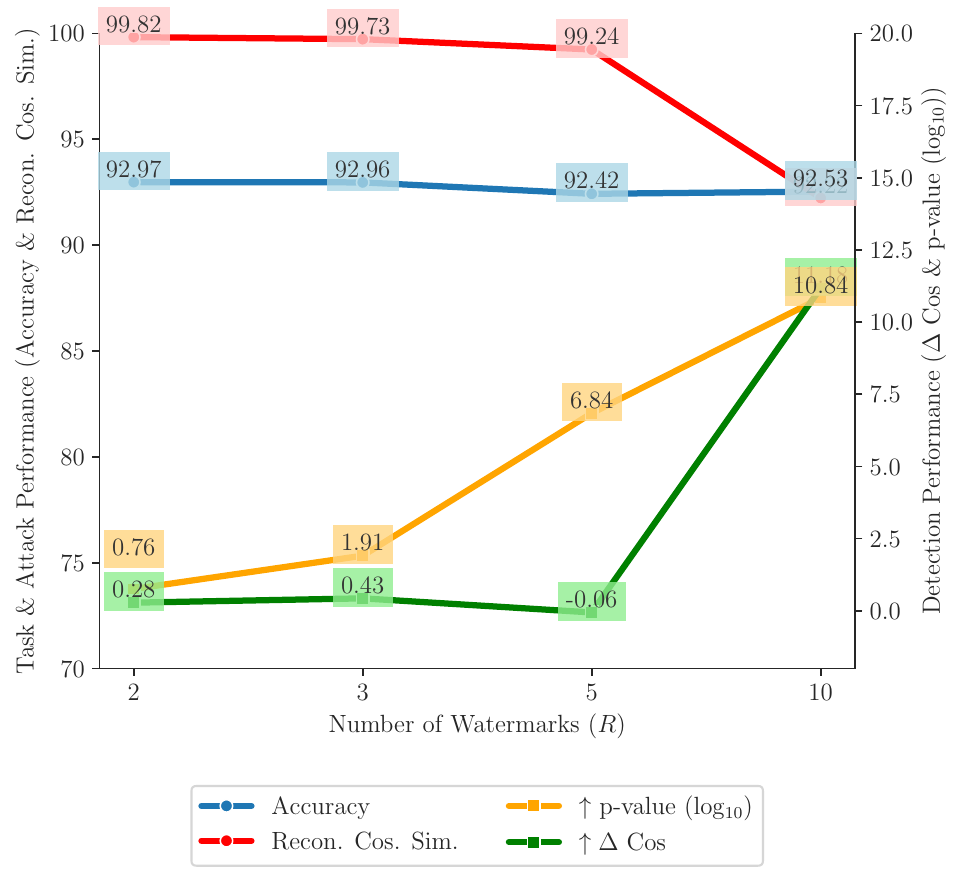}
        \caption{\agnews}
    \end{subfigure}
    \begin{subfigure}{0.22\textwidth}
        \centering
        \includegraphics[width=\linewidth,keepaspectratio]{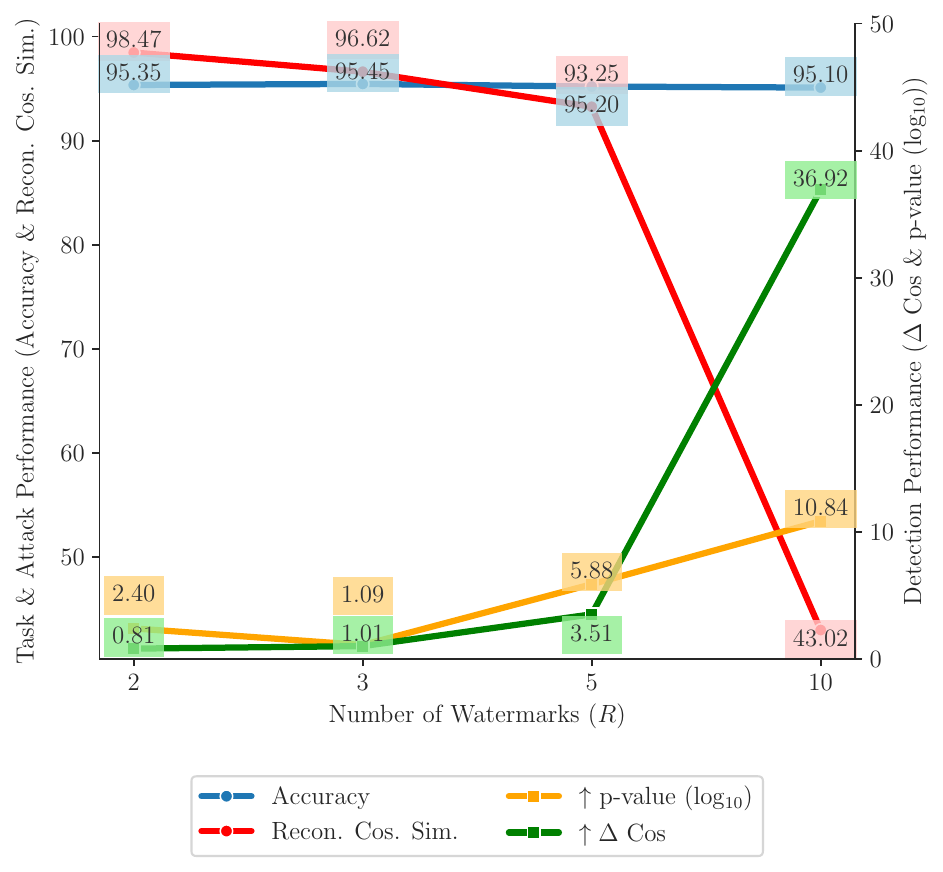}
        \caption{\enron}
    \end{subfigure}
    \caption{The impact of number of watermarks ($R$) in \ourdefence GS extension against \ourattack for different datasets. Same trend as  \ref{fig:num-watermarks-warden-cse}, but stronger metrics.}
    \label{fig:CSE-on-GSO}
\end{figure*}

In this subsection, we present the results (\reffig{fig:num-watermarks-warden}-\ref{fig:CSE-on-GSO}) for all the remaining datasets discussed in \refsec{sec:warden-exps}.

\begin{figure*}[h]
    \centering
    \begin{subfigure}{0.23\textwidth}
        \centering
        \includegraphics[width=\linewidth]{asset/imgs/ablation-GSO-defence/sst2_v2.pdf}
        \caption{\sst}
    \end{subfigure}
    \begin{subfigure}{0.23\textwidth}
        \centering
        \includegraphics[width=\linewidth]{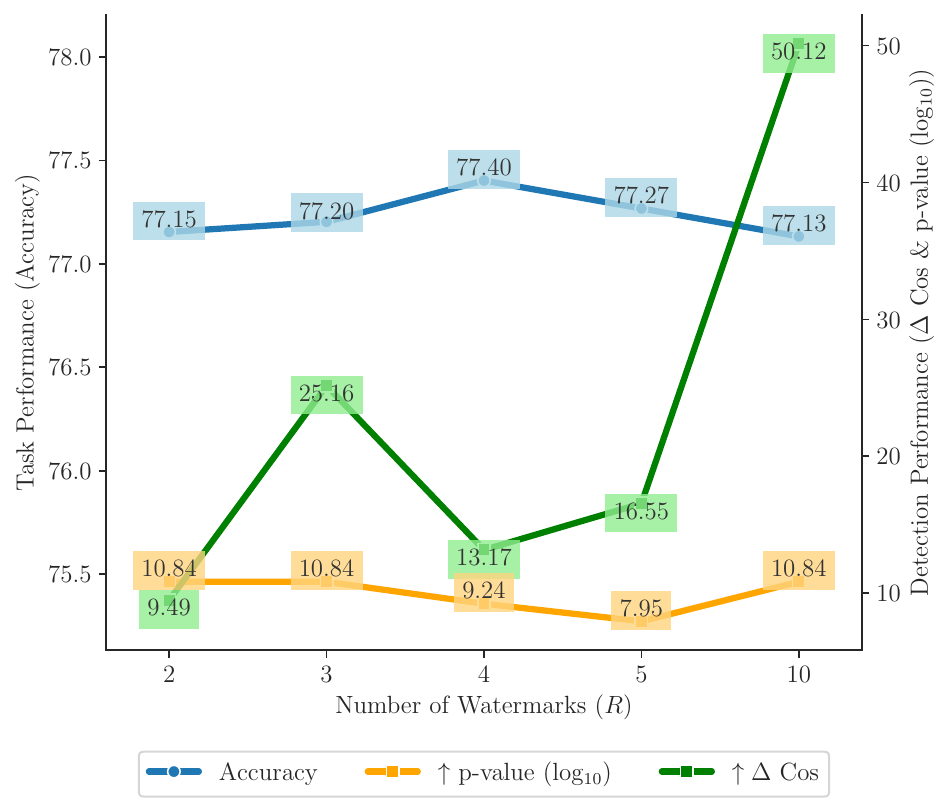}
        \caption{\mind}
    \end{subfigure}
    \begin{subfigure}{0.23\textwidth}
        \centering
        \includegraphics[width=\linewidth,keepaspectratio]{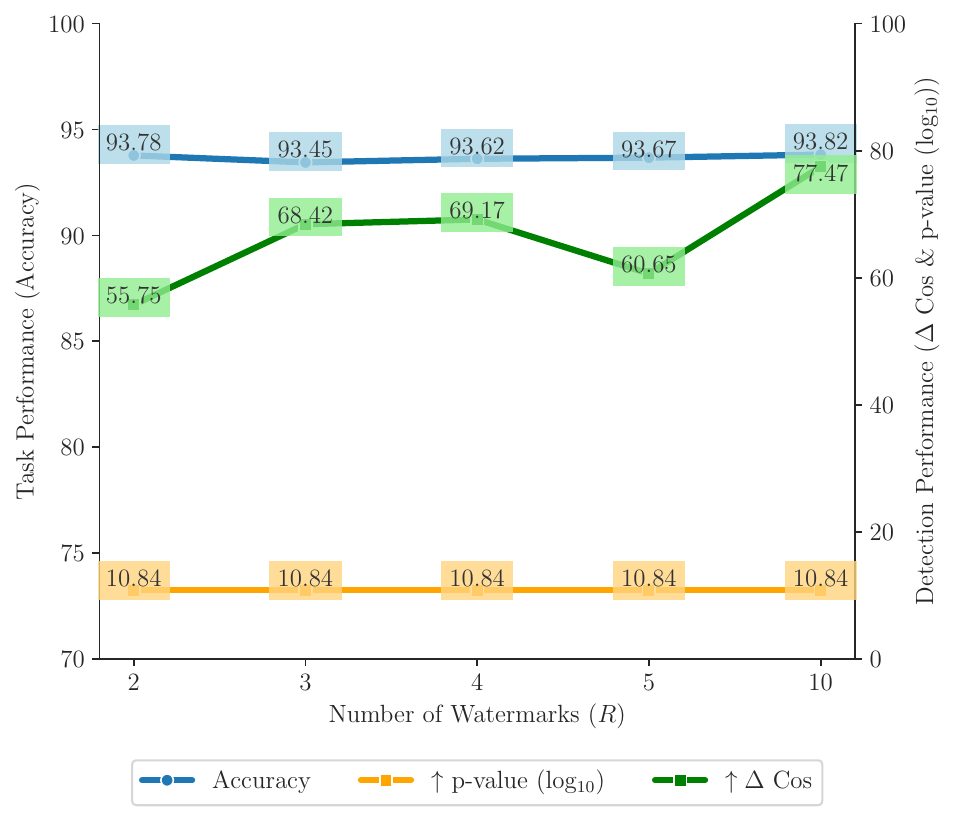}
        \caption{\agnews}
    \end{subfigure}
    \begin{subfigure}{0.23\textwidth}
        \centering
        \includegraphics[width=\linewidth,keepaspectratio]{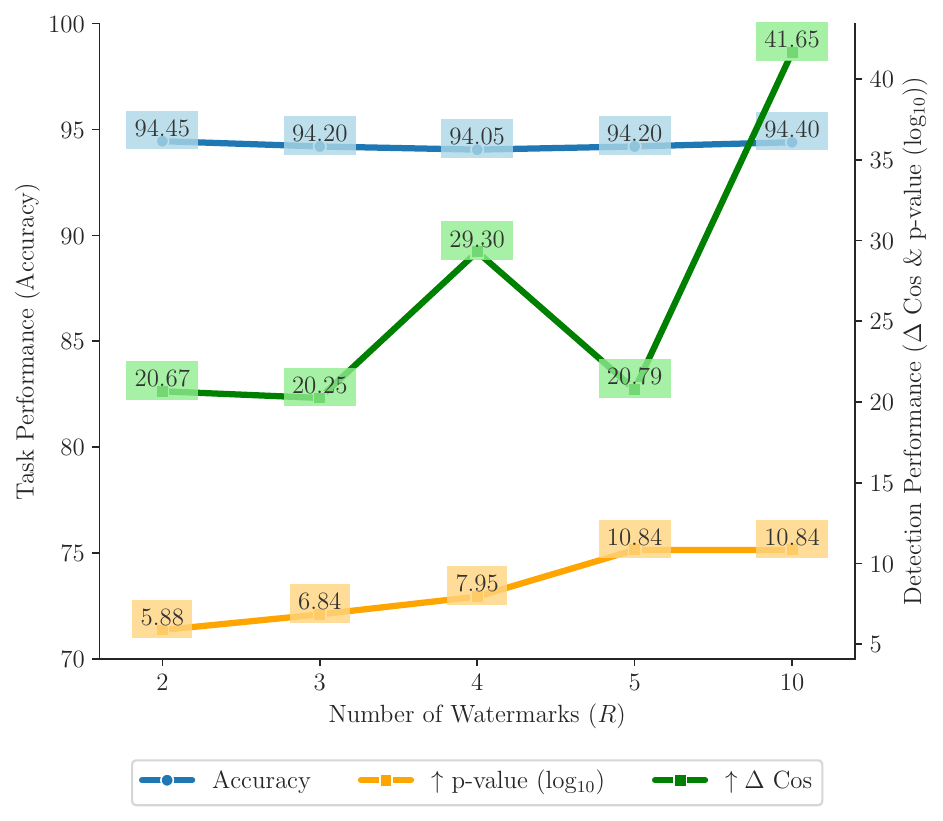}
        \caption{\enron}
    \end{subfigure}
    \caption{The impact of number of watermarks ($R$) in \ourdefence GS extension for remaining datasets. Same trend as \reffig{fig:num-watermarks-warden}, but stronger metrics.}
\end{figure*}

\subsection{Number of Principal Components ($K$) in \ourattack}
\label{app:warden-cse-pca-ablation}

\begin{figure*}[h]
    \centering
    \begin{subfigure}{0.22\textwidth}
        \includegraphics[width=\columnwidth,keepaspectratio]{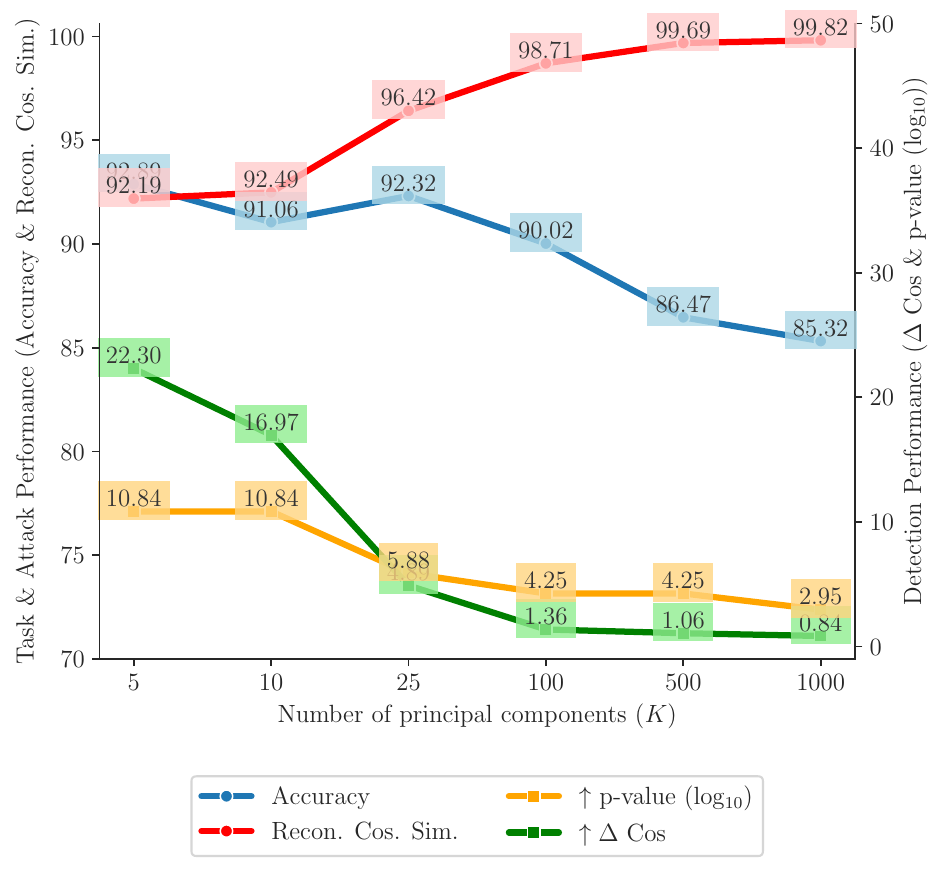}
        \caption{\sst}
    \end{subfigure}
    \begin{subfigure}{0.22\textwidth}
        \centering
        \includegraphics[width=\linewidth]{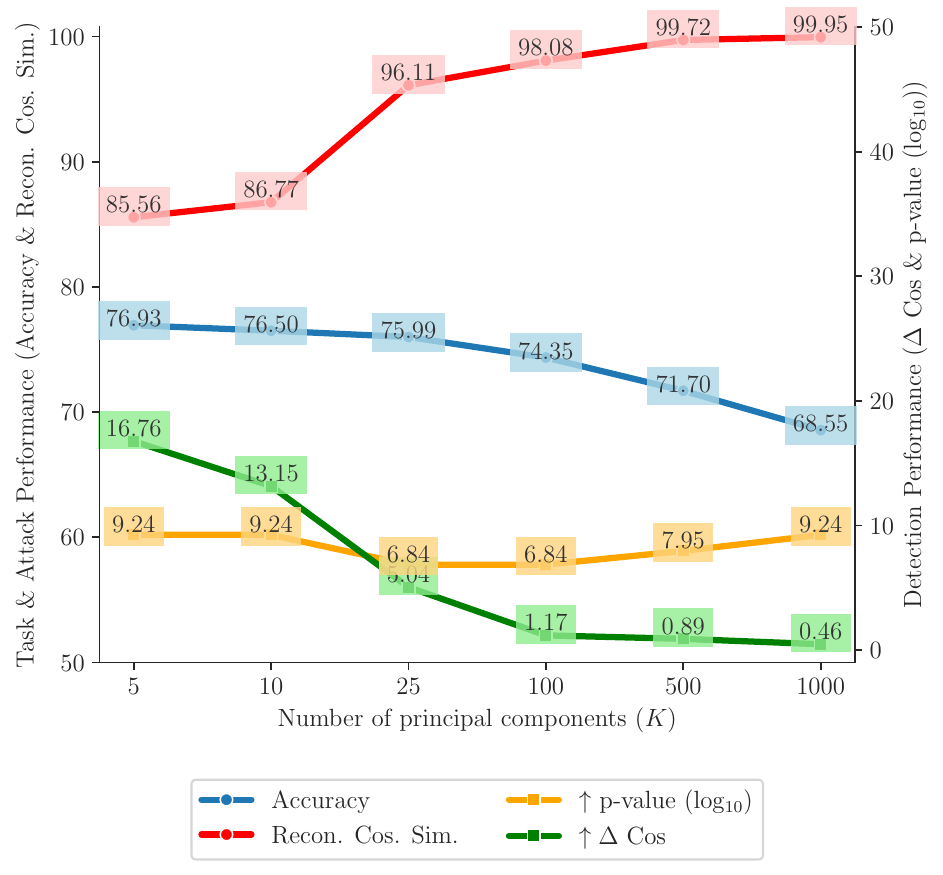}
        \caption{\mind}
    \end{subfigure}
    \begin{subfigure}{0.22\textwidth}
        \centering
        \includegraphics[width=\linewidth,keepaspectratio]{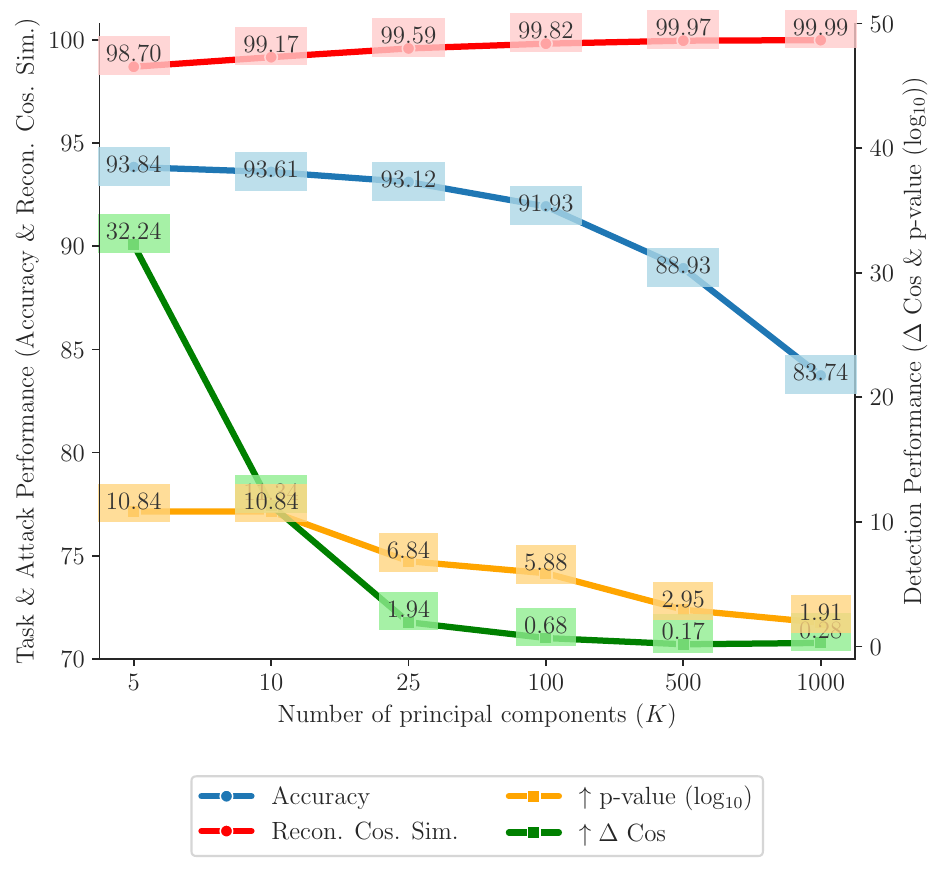}
        \caption{\agnews}
    \end{subfigure}
    \begin{subfigure}{0.22\textwidth}
        \centering
        \includegraphics[width=\linewidth,keepaspectratio]{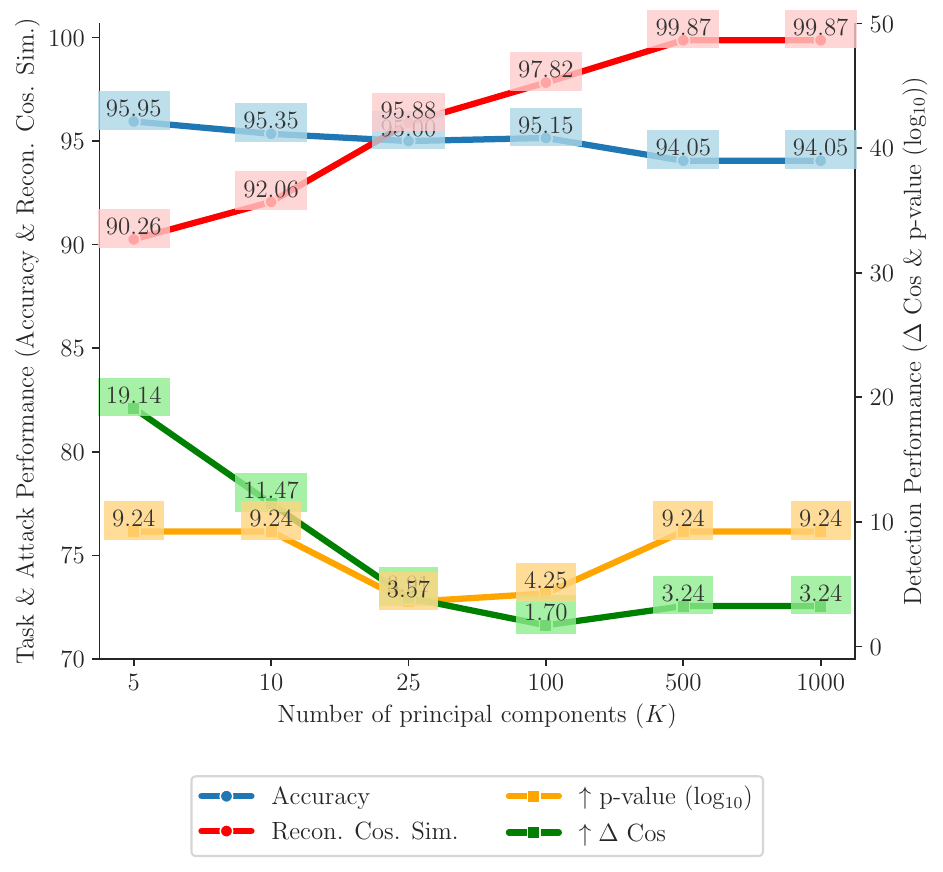}
        \caption{\enron}
    \end{subfigure}
    \caption{Against \ourdefence, the impact of number of principal components ($K$) in \ourattack for different datasets.}
    \label{fig:warden-pc-num}
\end{figure*}
Because we augment more target embeddings in \ourdefence, default configurations of \ourattack may not be adequate. We tweak the CSE attack by increasing the principal components eliminated, as shown in \reffig{fig:warden-pc-num}. We note that for a higher number of principal components, \ourattack is effective in some setups, as we recover more of the target embeddings. However, one thing to note is that the downstream metrics are poor if we eliminate a large number of principal components, undermining the attacker's objective. 

\subsection{Access to Target Embeddings}
\label{appendix:warden-remove-target-embs}
\begin{figure*}[h]
    \centering
    \begin{subfigure}{0.22\textwidth}
        \centering    \includegraphics[width=\linewidth,keepaspectratio]{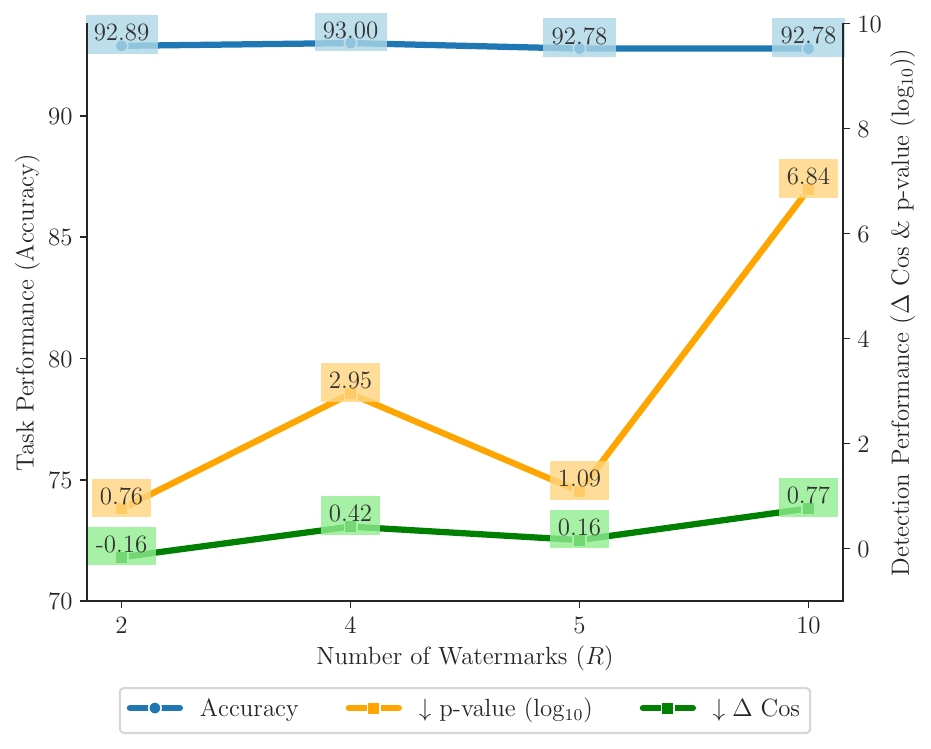}
    \caption{\sst}
    \end{subfigure}
    \begin{subfigure}{0.22\textwidth}
        \centering
        \includegraphics[width=\linewidth]{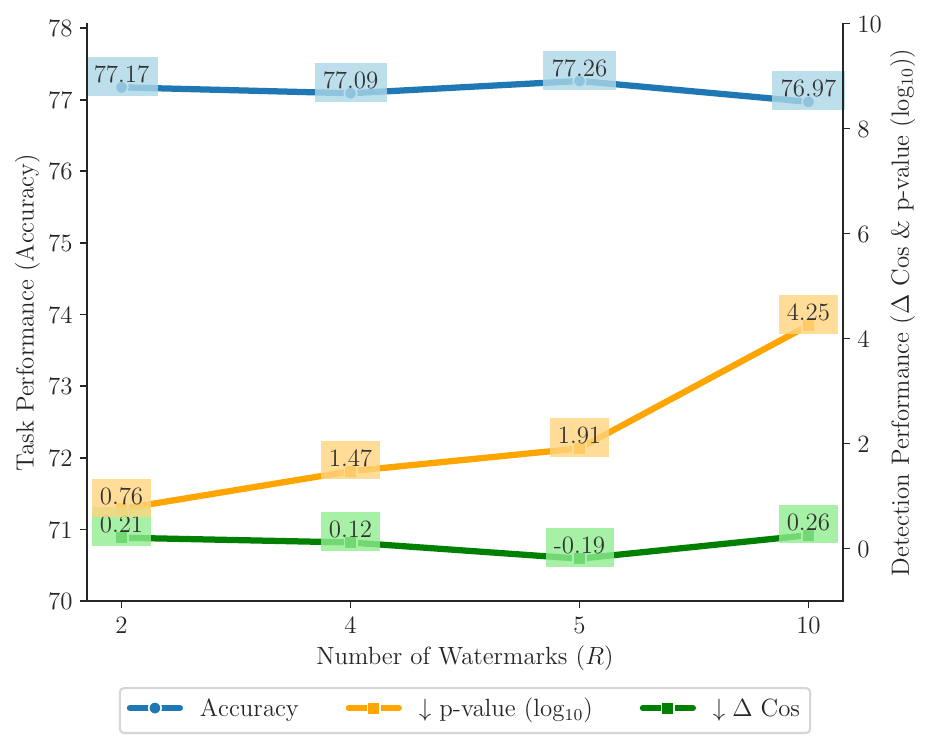}
        \caption{\mind}
        
    \end{subfigure}
    \begin{subfigure}{0.22\textwidth}
        \centering
        \includegraphics[width=\linewidth,keepaspectratio]{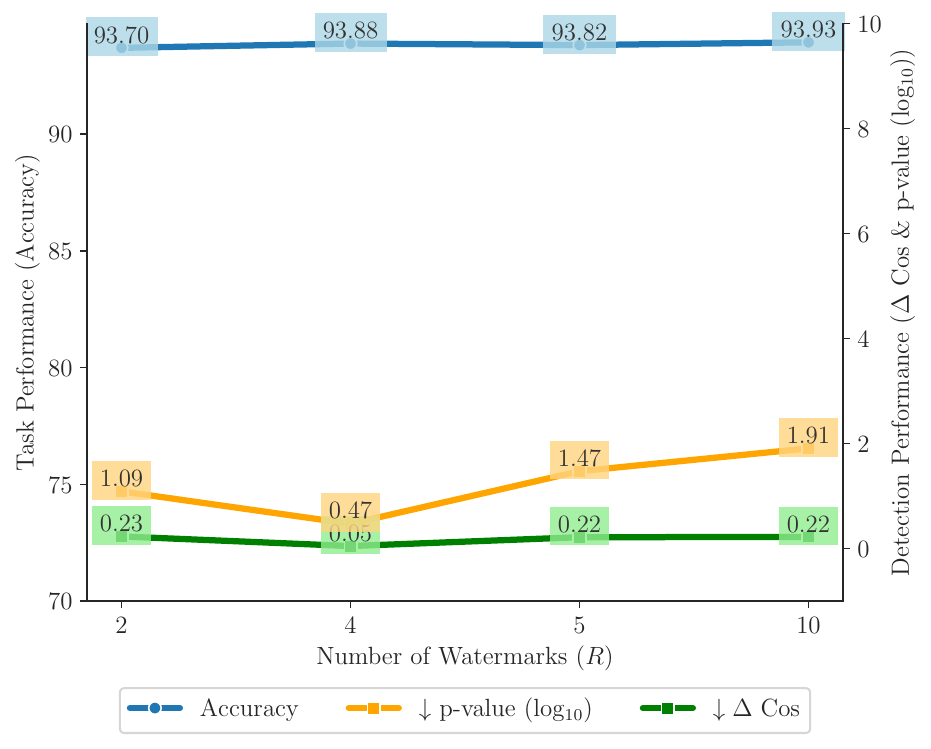}
        \caption{\agnews}
    \end{subfigure}
    \begin{subfigure}{0.22\textwidth}
        \centering
        \includegraphics[width=\linewidth,keepaspectratio]{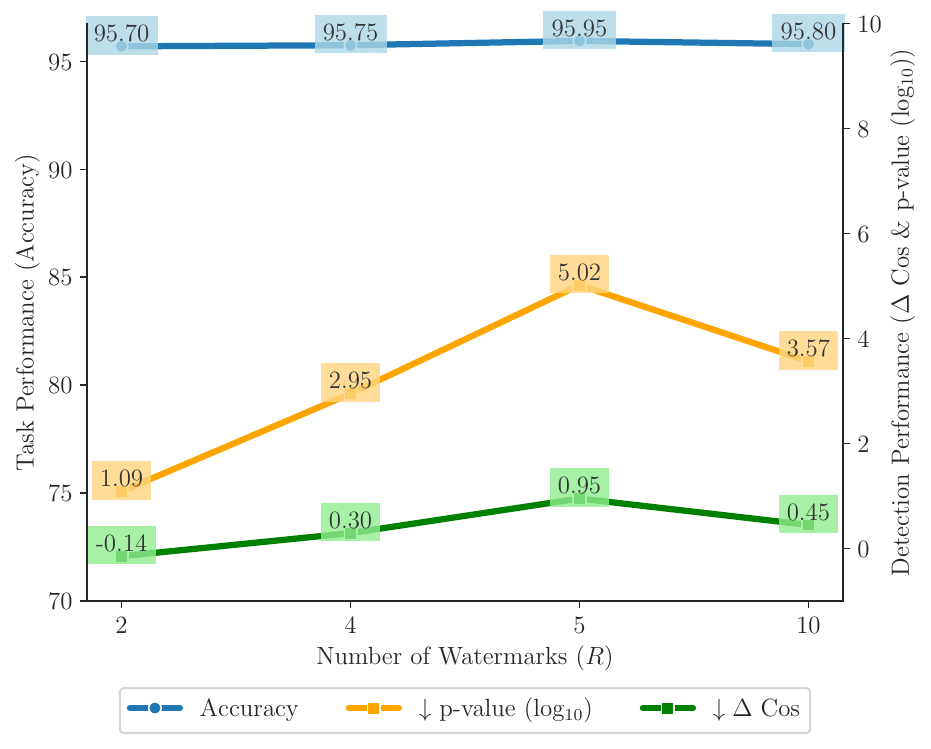}
        \caption{\enron}
    \end{subfigure}
    \caption{\ourdefence detection performance when secret watermarks ($\mW$) are eliminated for different datasets.}
    \label{fig:Warden-remove-target}
\end{figure*}

In the unlikely event that an attacker has access to all the target embeddings, it should be possible to bypass \ourdefence. We replicate this scenario and apply the elimination step of the \ourattack attack using these embeddings. As expected, in such a case, we can circumvent \ourdefence (see \reffig{fig:Warden-remove-target}).

\begin{figure*}
    \centering
    \begin{subfigure}{0.22\textwidth}
        \centering
        \includegraphics[width=\linewidth]{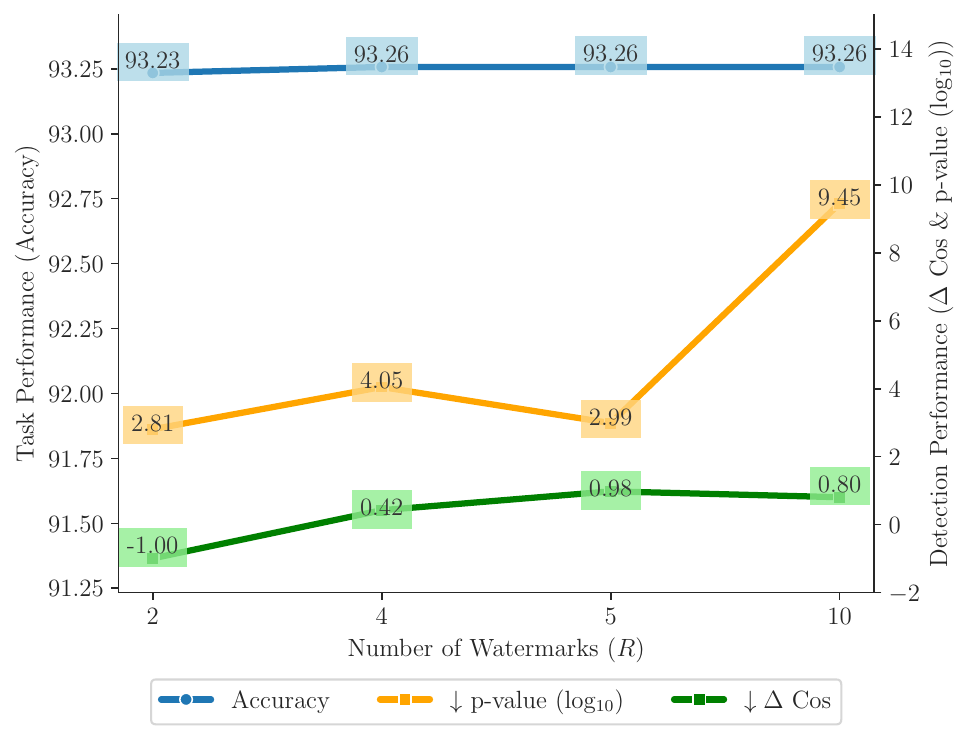}
        \caption{\sst}
    \end{subfigure}
    \begin{subfigure}{0.22\textwidth}
        \centering
        \includegraphics[width=\linewidth]{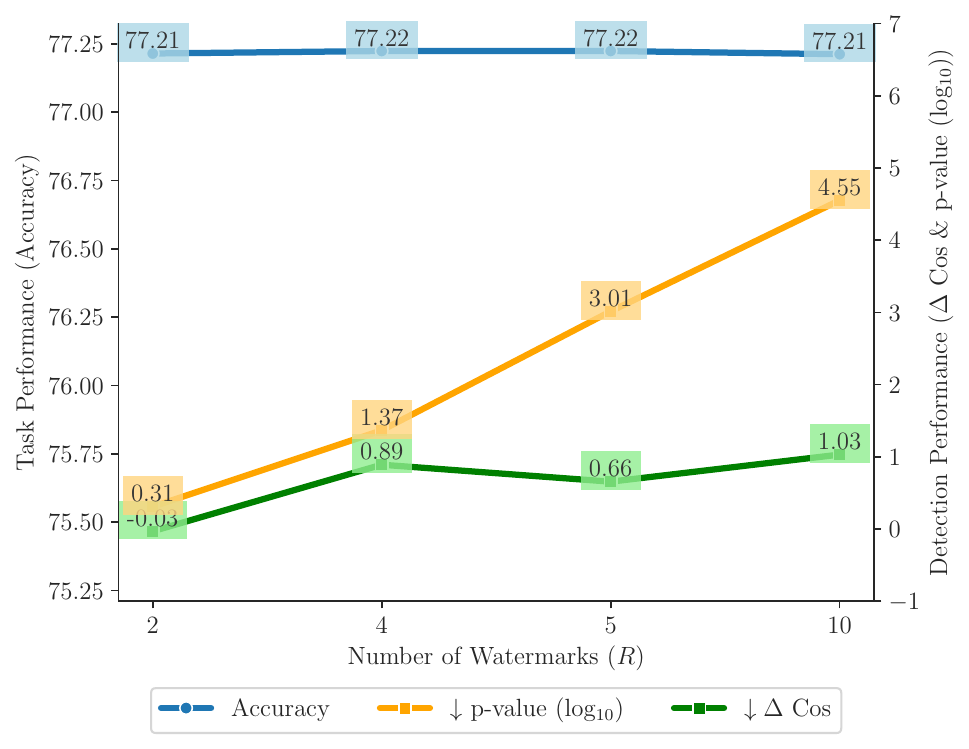}
        \caption{\mind}
    \end{subfigure}
    \begin{subfigure}{0.22\textwidth}
        \centering
        \includegraphics[width=\linewidth,keepaspectratio]{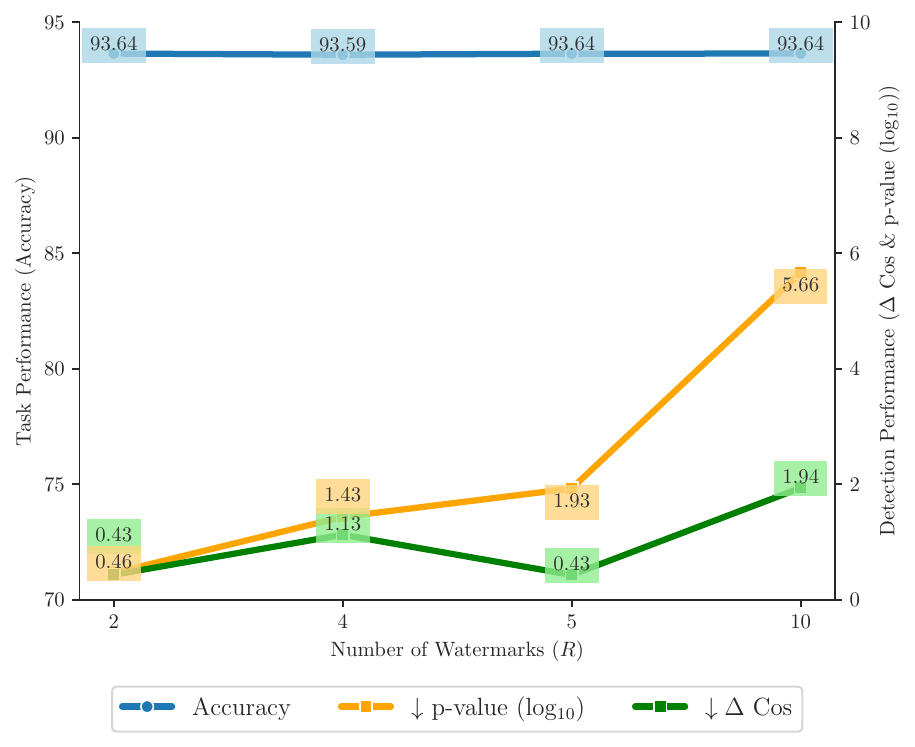}
        \caption{\agnews}
    \end{subfigure}
    \begin{subfigure}{0.22\textwidth}
        \centering
        \includegraphics[width=\linewidth,keepaspectratio]{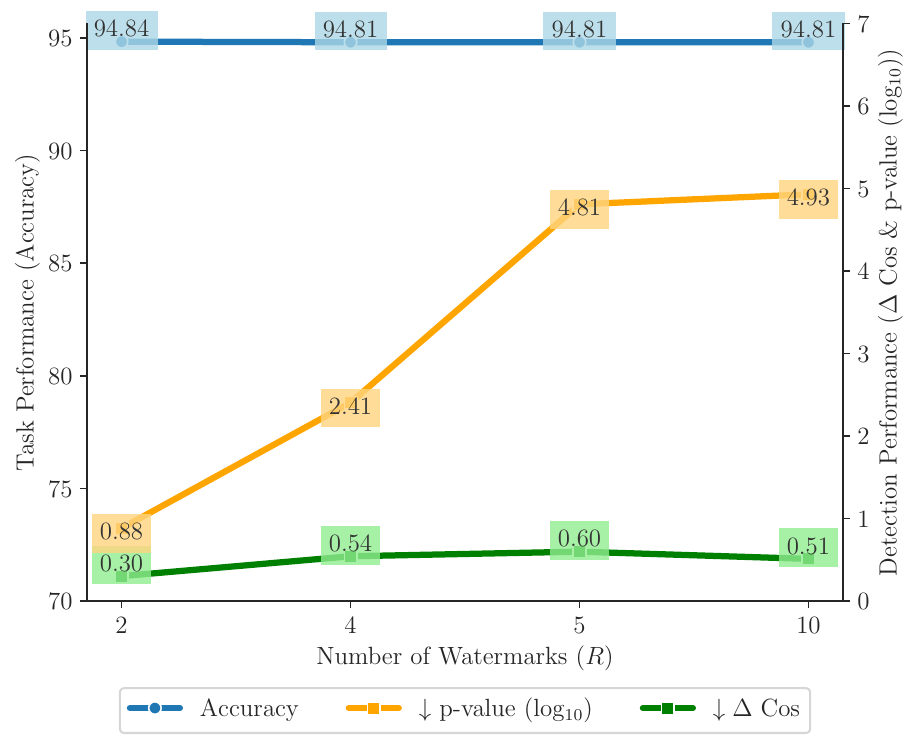}
        \caption{\enron}
    \end{subfigure}
    \caption{\ourdefence detection performance on a non-watermarked victim model for different datasets for different numbers of watermarks ($R$).}
    \label{fig:warden-non-watermarked}
\end{figure*}

\end{document}